Progress during the NOPP Wave Model Improvement Program

Donald T. Resio
University of North Florida

Charles L. Vincent
University of Miami

Hendrik L. Tolman
National Oceanic & Atmospheric Administration

Arun Chawla[1]
National Oceanic & Atmospheric Administration

W. Erick Rogers
Naval Research Laboratory

Fabrice Ardhuin
Ifremer

Alexander Babanin
University of Melbourne

Michael L. Banner
The University of New South Wales

James M. Kaihatu
Texas A& M University

Alexander Sheremet
University of Florida

William Perrie
Bedford Institute of Oceanography

---

[1] Corresponding Author: Arun.Chawla@noaa.gov

J. Henrique Alves
National Oceanic & Atmospheric Administration

Russel P. Morison
The University of New South Wales

Tim T. Janssen
Spoondrift Technologies Inc.

Pieter Smidt
Spoondrift Technologies Inc.

Jeff Hanson
WaveForce Technologies

Vladimir E. Zakharov
University of Arizona

Andre Pushkarev
Waves and Solitons, LLC

**Abstract**

This paper reviews the research activities that were carried out under the auspices of the National Ocean Partnership Program (NOPP) to advance research in wind wave modeling and transfer maturing technologies into operational community models. Primary focus of research activities that were funded under this program was to improve the source terms associated with deep water wind waves with a secondary focus on shallow water processes. While the focus has been on developing capabilities for stochastic (phase averaged) models, some of the research work reported here also touches on phase resolved models as well as updates that are needed to the classical stochastic equations to be applicable in shallow water conditions. The primary focus is on the development of new source terms to account for wave generation, dissipation and non-linear wave – wave interactions. A direct result of this program has been the development of new physics packages in operational wave models that have improved forecast skill from 30 – 50%. Since this is an overview paper summarizing all the activities that were undertaken under this program, only the major results are presented here. The readers are directed to other publications for more details. The paper ends with a discussion of the remaining major challenges in wind wave modeling, from the larger open ocean scales to the smaller coastal domains.

# 1.0 Introduction

In 2010 the Office of Naval Research, the National Weather Service, the US Army Corps of Engineers and the Bureau of Ocean Energy Management recognized a common need to update and enhance the operational wind wave models used by their respective agencies to perform forecasts, hindcasts and engineering analyses required for their missions. These agencies joined under the auspices of the National Ocean Partnership Program (NOPP) to support wave research by investigators from Academia, Industry and Government (Table 1) in what has become known as the NOPP Wave Research Program with the goal of transferring maturing technology into the Federal operational models. Tolman et al. (2013) provides an overview of the state of the art as well as the plans at the start of the program. The agencies recognized that they all had models customized to various approaches to the solution of the Radiative Transfer Equation (RTE) of Hasselmann (1962). Primary focus of the research was directed at deep-water waves with a secondary focus on improvements to shallow-water applications. This paper will review the results obtained and outline research challenges remaining.

With respect to the deep-water spectral modeling portion of this paper, it should be recognized that this paper is intended primarily to be an overview of progress on an operational model, although a limited set of supplemental investigations of deep-water source terms not executed in basin-scale WW3 testing were included in NOPP and will also be included here. An *a priori* modeling framework always serves as a constraint on what can be accomplished within a single project. Hence, there is no intent to represent this progress as the only relevant modeling approach available today; however, the authors feel that substantial progress made in many areas, some that were able to be incorporated into the WAVEWATCH III, which was used as the basic modeling system, and some that were not. There is no suggestion intended here that subsequent research is not needed, and the last part of the paper will provide a discussion of future pathways for such research.

In the models used by the agencies, the RTE paradigm for model formulation consisted of a series of energy source or sink terms solved on a grid allowing propagation of wave energy. This wave computation code was integrated into a significant infrastructure that handled information movement in and out of the code or for its interpretation. The approach for the program in deep water placed emphasis primarily on source/sink term development such that new versions could simply be implemented within their current models. The investigators transitioning their work into operational models accepted the offer of NWS to use their WAVEWATCH III wave modeling framework (hereafter WW3) as an integrating platform to allow consistent testing and evaluation of source terms within a common framework. NWS now considers WW3 to be a community model, while SWAN is an open-source model for shallow-water models. The goal was to allow potential users the options of using the results in WW3, or SWAN or in their own model. The more than 3000[2] users of WW3 and SWAN will have immediate access to the results.

The premise of quasi-homogeneous and near-Gaussian wave fields on which the RTE is based does not generally apply to wave propagation and source terms in shallower coastal areas. Efforts were directed to account for shallow-water effects (e.g. inhomogeneity and nonlinearity) either by 1) the development of first-principle generalizations of the RTE, 2) improvements of source terms and parameterizations, or 3) development of phase resolving approaches. These investigate improvements to the RTE framework and improve understanding of complex coastal processes. Where possible, the shallow water improvements would be incorporated into the operational suite.

This paper provides an overview of the principal results from the many investigators. Given the number of contributions to this work the complexity of the problems addressed, research results can only be presented in a somewhat summarized form here, with specific details left to referenced publications.

## 2.0 Spectral Wave Modeling

For any point with position **x** and time $t$ on the ocean surface, the wave energy is decomposed in spectral space, normalized such that the sum of the spectral density $F$ over all wave numbers $k$ is the surface elevation variance, obtained by averaging over the wave phases, at **x** and $t$. The spectral density $F$, evolves due to propagation and the effects of resonant nonlinear interactions, wind input, and dissipation (Hasselmann 1962, 1963a, 1963b):

$$\frac{\partial F}{\partial t} + \mathbf{c_x} \cdot \nabla_\mathbf{x} F + \mathbf{c_k} \cdot \nabla_\mathbf{k} F = S_{in} + S_{nl} + S_{ds} \quad (1)$$

The left side of Equation 1 is known as the Radiative Transfer Equation (RTE), where $\mathbf{c_x}$ and $\mathbf{c_k}$ are the velocities in physical and spectral space, with $\nabla_\mathbf{x}$ and $\nabla_\mathbf{k}$ the corresponding 2-dimensional gradient operators. The source terms on the right represent energy redistribution due to nonlinear interactions ($S_{nl}$), the net input (or loss) of energy due to the wind ($S_{in}$) and wave dissipation ($S_{ds}$), the latter generally dominated by wave breaking. Most applications for marine meteorology use an explicit estimation of $S_{nl}$, as is available in models such as WAM (WAM Development Group 1988), SWAN (Booij, et al 1999) or WAVEWATCH III (Tolman 1998, WAVEWATCH III Development group, 2016, hereinafter WW3) as opposed to earlier parameterizations (e.g. Barnett, 1968; Ewing 1971; Hasselmann, et al., 1975).

In presence of significant currents, it is convenient to avoid an extra source term for the interactions of wave and current energy (Phillips 1977), by rewriting the RTE as an action balance equation (e.g. Tolman et al., 2013),

---
[2] Estimate based on requests

$$\frac{\partial N}{\partial t}+\nabla_{\mathbf{x}}\left[\left(\mathbf{c_x}+\mathbf{U}\right)N\right]+\nabla_{\mathbf{k}}\cdot\left(\mathbf{c_k}N\right)=\sigma^{-1}\left(S_{in}+S_{nl}+S_{ds}\right) \tag{2}$$

where $U$ *is* the horizontal advection velocity for the wave action. In the case of depth-uniform currents, $U$ is independent of $k$ and equal to the mean current (Andrews and McIntyre 1978). $N=(F/\sigma)$ is the spectral density of wave action, and $\sigma$ is the intrinsic frequency that would be measured in a reference frame moving at the speed $U$ the advection velocity, which is independent of $k$ and equal to the mean current in the case of depth-uniform currents (Andrews and McIntyre 1978).

The evolution of wave models toward detailed, balance-form modeling as being done in third-generation wave models, was motivated by the need to represent the physics of wave generation, propagation and dissipation in a single, unified framework (SWAMP Group, 1985). In practice, recent parameterizations correspond to a re-definition of the source term as atmosphere-wave ($S_{in} \triangleq S_{atm}$), wave-wave (unchanged) and wave-ocean ($S_{ds} \triangleq S_{oc}$) terms is required by the definition of fluxes in coupled models (Ardhuin et al. 2010). This clarifies that swell dissipation due to friction at the air-sea interface and negative cospectral energy transfers between pressure perturbations and the vertical motions of the water surface are included in $S_{in}$, whereas the part due to interaction with ocean turbulence is included in $S_{ds}$. Here we will not use the term "physics" to refer to a particular set of source terms being tested. Instead, we will use the term "source term parameterization," which emphasizes that our understanding of the underlying physics is still incomplete, and that parameterized approximations within models undergo continuous improvement to improve the state of the art (e.g. WAVEWATCH III Development Group, 2016).

Although the scope of the work reported here is quite broad, emphasizing processes affecting global scale modeling to smaller-scale coastal wave dynamics, it by no means addresses all aspects of wave modeling. For instance, we considered partial blocking by icebergs (Ardhuin et al. 2011), but not wave propagation through ice fields, or ice effects on wave generation, the latter topics are more extensively studied in several ongoing research initiatives (e.g. Thomson et al. 2018). Closer to shore, we consider sandy or muddy seafloors and the presence of vegetation, but not the effects of wave propagation over reefs. On the other hand, the development and testing of improved source terms suitable for implementation in stochastic wave models, and the development of new propagation models for in-homogeneous and non-Gaussian wave statistics, should eventually contribute to a more comprehensive wave model.

Deep-water and shallow-water wave dynamics are generally quite distinct, often posing very different requirements on grid resolution, source term parameterizations, and even transport models (e.g. Roland and Ardhuin 2014). In particular the bottom topography and nature play a particular role, and the very short space and time scale of evolution can be better resolved with specific numerical methods. Because of these distinct dynamics, the remainder of the paper is divided into three parts: deep-water, shallow-water and an overall discussion of challenges.

## 2.1 Deep Water Modeling Development and Testing

Developments reported here can be partitioned into contributions of three different types of testing and development:

1) provision of complete and new source-term sets in publicly available codes used operationally (WW3 or SWAN), with thorough testing at local to global scales;

2) development of specific source terms which are designed to replace individual source terms within these models, but have yet not been tested at a basin-wide scale; and

3) development of improved theoretical foundations for source-term physics.

In all cases, the work is reported in peer-reviewed publications, briefly reviewed in subsequent sections. All three are important to the field of wave forecasting and hindcasting given the many source terms assumptions and modeling approximations inherent in the complexity of the coupled ocean-atmosphere coupling responsible for wave generation and dissipation.

Here we review in particular the development of two new deep-water parameterizations which underwent detailed validation at global scale and have been implemented in most forecasting centers. They have been validated against a wide range of measurements (e.g. Rascle and Ardhuin 2013, Zieger et al. 2015) and compared to each other (Stopa et al. 2016) and to previously existing parameterizations (Rascle and Ardhuin 2013, Stopa et al. 2016). After recalling their main features we will give some examples of their performance, using the same numerical integration schemes. The different pairs of parameterizations for input and dissipation discussed here are named following the notation of "switch" options in WW3. These are ST1 (WAMDI 1988), ST2 (Tolman and Chalikov 1996, Tolman 2002), ST3 (Janssen et al. 1994, Bidlot 2012) and the two parameterizations developed and further tested, ST4 (Ardhuin et al. 2010, Rascle and Ardhuin 2013), and ST6 (Rogers et al. 2012, Zieger et al. 2015). Another parameterization, ST5 based on Banner and Morison (2010) has been essentially developed with the full integral form for $S_{nl}$, but it has only undergone limited testing for real events.

The latest evolutions of ST4 and ST6 are described in the latest WW3 manual (WAVEWATCH III Development Group, 2016, 2018). Table 2 summarizes the main features of these parameterizations. The ST1 and ST3 parameterizations used a mean steepness to parameterize the wave dissipation all across the spectrum, with no distinction of physical processes (breaking, air-sea friction, wave-turbulence interaction), and a prescribed shape of $S_{ds}$ proportional to $k$ and $(ak + bk^2)$ and to the directional wave spectrum. These two parameterizations have undesirable and unrealistic side effects: a swell generally produces a strong reduction in wind sea dissipation (van Vledder 1999; Ardhuin et al. 2007), and a wind sea produces a strong increase in swell dissipation (Rogers et al. 2003), as illustrated in figure 1. The representation of swell dissipation using steepness-limited breaking is particularly implausible. Further, the balance in the tail of the spectrum is not well reproduced and requires a "diagnostic tail" making it impossible to reproduce slope statistics (e.g. Munk 2009). The problems with swell dissipation was one of the main reasons for the development of ST2, which heuristically separated a low frequency and high frequency dissipation to avoid spurious windsea-swell interactions, and used an adjustment of the negative part of $S_{in}$ to reproduce swell heights in the Pacific (Tolman 2002). Although the evolution from ST3 and ST2 to ST4 and ST6 was largely motivated by a new paradigm for wave dissipation (Phillips 1985, Banner et al. 2000), the difference in the generation source term $S_{in}$ should not be underestimated. Figure 1C and 1D show estimated effects of swell on the drag coefficient $C_D$ which is used to estimate the friction velocity from which $S_{in}$ is computed. In ST3 and ST4, a swell generally reduces the drag, which then reduces $S_{in}$, but the swell has no impact on ST4 dissipation whereas it strongly reduces dissipation in ST3. Because ST6 employs an empirical drag coefficient that depends only on wind speed, such effect is absent and swell effect is minimal on windsea growth.

The later parameterizations ST4 and ST6 build on these ideas, but use a completely different paradigm for the wave breaking term. Following Phillips (1984, 1985) and Banner et al. (2000) a wave breaking probability and associated dissipation rate are estimated, as a function of a wave steepness that varies across the spectrum. ST4 and ST6 differ in the ways this steepness is estimated from a spectral bandwidth or locally (we also note the smoothing in Banner and Morison 2010), and how the dissipation of waves is redistributed across the spectrum, with several options available in ST4 (Filipot and Ardhuin 2012, Leckler et al. 2013). An important result is that the wave breaking does not appear to be a function of the near-local steepness only, and appears to be significantly influenced also by longer waves (Young and Babanin 2006). Hence ST4 and ST6 have defined heuristic added dissipation due to the effect of longer waves presumably via their breaking (Ardhuin et al. 2010, Babanin et al. 2010). These wave breaking parameterizations are rather well constrained for the breaking probabilities of dominant waves, and more recent observations of dissipation rates (e.g. Sutherland et al. 2015) may be used to further refine the parameterizations (e.g. Romero et al. 2012). Also, following the idea of Phillips (1984), the investigation of an out-of-equilibrium situation with current gradients can be an ideal, but generally complex, situation for testing the different parameterizations with very large differences between model results (e.g. Ardhuin et al. 2012).

These parameterizations are still insufficient to obtain the observed spectral shapes with $F(\boldsymbol{k}) \sim |\boldsymbol{k}|^{-4}$ for |k| larger than twice the windsea peak (Leckler et al. 2015, Peureux et al. 2018), giving spectral tails closer to $|\boldsymbol{k}|^{-3.7}$ (Zieger et al. 2015). As a result, when investigating surface slope statistics, the default settings of ST4 and ST6 still require a diagnostic tail when models are run with a maximum frequency above 0.6 Hz. That defect is the topic of ongoing research, focusing in particular on short wave modulation by long waves which gives promising results (Peureux 2018), but requires a complete retuning of the model.

Another important evolution from ST2 to ST4 and ST6, is the definition of a swell dissipation that is calibrated with measured non-linear swell dissipation rates in the far field of storms (Ardhuin et al. 2009). The parameterization adapted by Rascle and Ardhuin (2013) uses a smooth transition from a laminar to a turbulent air-side boundary layer that is justified by the Rayleigh distribution of wave heights (Stopa et al. 2016b). The swell dissipation formulation in Zieger et al. (2015) gives similar values. These parameterizations behave very differently from ST2 or ST3 in the wind sea to swell transition regime, for which very little data is available. The future satellite missions such as CFOSAT (Hauser et al. 2017) should contribute to fill this gap.

## 2.2 Basin-Scale Model Tests

Comparisons of different wave models such as routinely performed under the Joint Commission on Marine Meteorology (JCOMM) wave model verification exercise are obscured by differences in wind forcing, data assimilation, grid resolution and other features (Bidlot et al. 2007). Here we report on already published validations at global scales using a relatively coarse grid with resolutions of 0.5 degree in longitude and latitude. One of the main difficulties in this type of wave hindcast, is to define a consistent, unbiased forcing field for testing. Indeed, wind speed underestimation can produce a very severe storm and swell underestimation (e.g. Caires and Sterl 2005). Detailed investigations have shown that ECMWF wind products tend to underestimate extreme wind speeds (Stopa et al. 2018, Pineau-Gillou et al. 2018) and for this reason we will discuss the comparison of results obtained with ST2, ST3, ST4 and ST6 already published by Stopa et al. (2016), using winds from the Climate Forecast System Reanalysis (CFSR, Saha et al. 2010). Further comparison with ST1 can be found in Roland and Ardhuin (2014). The other forcing fields used are sea ice concentrations from CFSR and iceberg concentrations from Tournadre et al. (2008, 2012). These comparisons reveal reasonable estimates for integrated spectral parameters in most open-ocean areas for all parameterizations. Starting with wave heights, Roland and Ardhuin (2014) and Stopa et al. (2016) showed that biases generally involve a lower bias on east coasts compared to west coasts, with most of the ocean between -20 and +20 cm compared to satellite altimeters which have average oceanic values of 2.2 to 2.3 meters (Zheng et al., 2015). Also, with the models adjusted to large basins, ST4 and ST6 generally give lower values than ST3 in enclosed seas like the Mediterranean. Differences between parameterizations are generally larger for the root mean square error, as shown in Figure 2, with a general reduction by 30 % in going from ST3 to ST4 or ST6, the error reduction is even larger compared to ST2. Errors with ST3 are particularly large in regions where both swell and young windsea are present, such as the gap wind regions of the Pacific coast of Mexico, or the east coasts off New England and Japan. We attribute this reduction in error to the more realistic decoupling of windsea and swell that is used here related to the saturation-based dissipation (e.g. Ardhuin et al. 2012) as was illustrated in Figure 1. Stopa et al. (2016) further showed the larger scatter index for the mean square slopes (mss) obtained with ST3 compared to ST6, ST2, and ST4. These mss diagnostics give a very strong constraint on the variability of the spectral tail, which is generally not realistic with ST3 as shown in Figure 1 (see also Ardhuin et al. 2010).

Other spectral parameters appear consistent with this general behavior, although comparisons are limited to the few locations where buoy data are available. This is reproduced in Figure 3. The large standard deviation obtained with ST3 for the mss, $m_2$ and $T_{m02}$ are associated to spurious variability of the high frequencies, which is also a problem with ST6 but only at the highest frequencies, and thus visible only in the mss. We also note that ST6 gives a more realistic standard deviation of Hs, close to the observed value, possibly associated to a stronger growth for young waves. The problem with high frequency energy level in ST6 has been corrected: in SWAN in 2014 (and more recently in WW3 (Liu et al., submitted).

The only directional parameter shown in Figure 3 is the directional spread, computed for the entire spectrum, for which ST2 to ST6 give comparably poor performance. This points to remaining deficiencies in the representation of spectral shapes for which swell dissipation plays an important role as detailed in Stopa et al. (2016). Here we particularly discuss the other source terms.

The details of the frequency-direction distribution of wind energy entering a spectrum and leaving the spectrum due to wave dissipation remain a relatively open question, with different theories and concepts adopted from different combinations of observations and theoretical arguments. Because it is not yet feasible to compute the full nonlinear integral in high resolution forecasting models, the DIA (Discrete Interaction Approximation, Hasselmann et al. 1985) is usually used to represent $S_{nl}$.

While the hindcasts of specific data sets are very encouraging in terms of the error characteristics, it is important to understand the differences between the source term sets in cases where differences are easier to interpret. To keep our comparison as simple and straightforward as possible, the results shown here are for a duration-limited test, assuming infinite fetch and a wind speed of 15 m/s. Figures 5 and 6 contain six panels, for snapshots at 3 hours of simulated time and 15 hours of simulated time, respectively. The upper left-hand panel in each shows the directionally integrated spectra produced by WW3 using source packages ST2, ST4, and ST6 and SWAN using ST6 source terms with 30-second and 180-second time steps.

After 3 hours of simulation time, the modeled spectra exhibit considerable variability, with the newer source term packages developed under the NOPP effort (ST4 and ST6) producing higher spectral values with lower peak frequencies. Whereas the WW3 results were observed to be relatively independent of time step, the SWAN results exhibited significant difference using different time steps with the result from a 30-second time step deviating markedly from the WW3 using ST4 and ST6 source term package. The 180-second results in this test are close to the spectrum produced by the older ST2 source term package, even though it used the ST6 source term package. The adjacent right-hand panel shows a log-log plot of these same spectra. The sensitivity to time step is discussed further below.

The left-hand panel in the second row provides the wind input source terms at the snapshot in time. The right-hand panel in this row shows the dissipation source term at this time. In the bottom row, the left-hand panel shows the nonlinear interaction source term and the right-hand panel shows the sum of all three source terms. In these comparisons it should be recognized that the source terms represent snapshots for a fixed time and not for a fixed spectrum, since the spectra have evolved differently over these three hours. The results using ST4 and ST6 show that the new wind inputs for young waves are considerably enhanced over the older ST2 source term package. It can also be seen that ST4 produces a broader wind input than ST6. The SWAN simulations using different time steps are similar in form to the ST6 WW3 results at this time, but shifted to a higher peak frequency, consistent with the shift in the location of the spectral peak seen in the upper right-hand panel. Comparisons of the dissipation source term show that the form of the dissipation is different in ST4 and ST6. Although the spectra for these two WW3 simulations are quite similar at this time, the dissipation in ST4 is larger and broader than that of ST6. The comparisons of the nonlinear transfer source term in the bottom row shows a pattern that, in general, appears consistent with the expected results from the DIA.

A noticeable feature of Figure 4 is the strong dependence of the SWAN result on time step size. The dependency is caused by the so-called limiter in SWAN, which limits wave growth to some fraction (usually 10%) of the spectral level of the parametric spectrum. The limiter follows that used in WAM Cycles 1 through 3 (see discussion in Tolman 2002). It is implemented to prevent fast oscillations in the tail. In WW3, the problem is addressed using dynamic time stepping (Tolman 1992); in that scheme, the time step is dynamically reduced to accommodate fast changes, which reduces the activity of the limiter. In SWAN, which does not have dynamic time-stepping, the limiter is active during periods of very fast wave growth. Figure 4 clearly falls into this category: here we are looking at the first three hours, starting from rest, under gale-force winds. However, after 15 hours (Figure 5), the impact of the limiter is apparently small. Further, it is our experience that in "realistic" simulations with SWAN, there are only minor differences in skill using SWAN with time steps sizes of, say, 5 and 10 minutes. This makes it difficult to justify use of the smaller time step size in an operational setting, where run-time is a primary concern. However, it is clear that there is room for improvement in the numerics of our operational codes (SWAN in this case), since the role of the limiter may be significant in some realistic cases. As we have shown here, the limiter impedes our ability to make clear comparison of the detailed behavior of source terms and could produce problems in applications to small coastal areas or basins. Interested readers are referred to Tolman (1992), Hersbach and Janssen (1999), Tolman (2002), and references therein.

The right-hand bottom panel shows the sum of all the source terms. With the exception of the SWAN simulation using 180-second time steps, the net source terms are small above about 1.3 times the frequency of the spectral peak. This re-emphasizes the point that the source terms in each source term package are balanced to achieve this end and that they are not intended to be used separately in different combinations.

The information in Figure 5 is similar to that in Figure 4 and shows the same set of comparisons as shown in Figure 4, except for 15 hours of simulation time rather than 3 hours. The modeled spectra exhibit considerably less variability at this stage of development, with the newer source term packages developed under the NOPP effort (ST4 and ST6) producing lower frequency spectral peaks than the older ST2 source-term package. The comparison of wind-input source terms shows the type of consistency that we might expect given the relatively similar spectra at this time. This panel clearly shows the difference between the wind input terms in ST4 and ST6, with the wind source in ST4 considerably more concentrated in the vicinity of the spectral peak than that in ST6, which has larger wind input in the tail. At this simulation time, both ST4 and ST6 produce lower wind inputs than ST2 near the peak. The differences in dissipation source terms do not match the expected differences needed to balance differences in the wind inputs, with the largest dissipation in ST4 rather than ST2. The nonlinear source terms all seem consistent with the DIA; and with the exception of SWAN with the 180-second time step, the sum of the source terms form a similar set in which the net changes are all similar and confined to the region near the spectral peak, with larger rates of energy gain on the low-frequency side of the peak and smaller rates of loss on the high-frequency side.

Stopa et al. (2016) clearly indicates that the ST4 and ST6 source terms are a significant improvement over the earlier source term sets based on a very comprehensive data set relevant to the global forecast problem. The levels of RMSE and skill score achieved might lead to the question as to what fraction of the error no longer lies in the wave model but in the representation of the wind field. However, analysis of the swell data clearly shows that while ST4 and ST6 significantly improve swell forecasts, room for improvement still appears possible. Given the relative importance of swell globally, typically dominating 75 to 95% of the time (Chen et al. 2002), further research in this area is warranted.

The improvement in skill represented by ST4 and ST6 has led to its adoption by many forecasting centers, starting with Meteo-France, NOAA/NCEP, FNMOC and NAVOCEANO in the U.S., the U.K. Met. Office and Envirnnement Canada, and probably many others. ECMWF is planning to change from ST3 to an adaptation of ST4 for use in the ECMWF wave model (J. Bidlot, personal communication). The adaptation is necessary because of ECMWF requirements for CPU time. This is a significant achievement of the research efforts. The improvements in skill has come at an increase in cost by 40 to 100% in ST4/WW3 and 25 to 50% in ST6/SWAN (Edwards et al. 2018), depending on the detailed model set-up, with the larger relative increase corresponding to models using first order advection schemes. In the case of ST4, the recent optimization performed at ECMWF roughly corresponds to an earlier version (TEST405) but these have generally been offset by increases in computer power. However, a point is reached in operational modeling in which much larger increases in computational resources are required to achieve only incremental model improvement.

The achievement of the ST4 and ST6 groups in improving the performance of the model to the degree shown provides significant improvement to forecast products. However, questions regarding the formulation of individual source terms indicate that more research is required to more fully understand the wind wave forecast problem. Two major changes have been made to the source term balance in WW3. First, the total integral of momentum flux into the wave field via the wind input source function was made consistent with our best knowledge of wave supported stress, either by using the approach of Janssen (1991) in ST3/ST4 or by applying empirical constraints in ST6 (Tsagareli et al., 2010); and second, swell decay within the model was defined and calibrated to be consistent with observations (Ardhuin et al., 2010; Stopa et al. 2016a). Of course, the advances cannot be wholly ascribed to the NOPP program. As noted, the wind input formulation of ST4 builds on prior work by Janssen (1991) and the separation of swell dissipation from whitecapping was already started by Tolman and Chalikov (1996). Taken as a whole, these works more firmly establish expectations for development of source functions for global models going forward.

## 2.3 Development of individual source terms for transfer into WAVEWATCH III

As noted in the previous section, constraints on the integrated momentum transfer have produced more consistency in the integrated growth rates achieved within different source term packages. Increased emphasis on coupled modeling and higher order spectral phenomena in shallow water make it important to include comparisons of modeled spectral shapes to observed spectral shape, which is very dependent on the details of the frequency-angle distribution of wind input and wave dissipation. An important topic is the degrees of freedom in comparison between modeled and measured directional spectra. Some spectral shape comparisons have been cast in very subjective terms. For example, directional spreading is evaluated in some published studies simply through visual comparison of two spectra side-by-side, while others used carefully defined quantitative metrics (see literature review by Rogers and Wang 2006). An important step toward improving our ability to quantify detailed spectral comparisons would be to identify general characteristics within observed directional wave spectra. Resio et al. (2016) lists a set of spectral characteristics which have been documented in multiple studies, which offers some useful guidance on this topic. An example is the "lobe ratio" used by Banner and Young (1994) which quantifies bimodal directional spreading in the tail. Since all source term sets have to be balanced to achieve a reasonable overall rate of energy gain and rate of shift in the peak frequency, metrics for this type of testing will likely be substantially more demanding than the metrics used in integrated-parameter testing.

Since the DIA remains the primary operational approximation for $S_{nl}$, the overall balance within the wind input and dissipation source terms has been calibrated to work in conjunction with the DIA. Recently, efforts have begun to examine the effect of switching to different versions of $S_{nl}$ on both the total integral balance and the detailed spectral shape. This is discussed in Section 3.

Continued development of ST5 (Banner and Morison, 2010) has also shown the need for changes in the wind-input and wave-dissipation terms in that model version in order to achieve consistency with observed growth rates and spectral shapes. A good example of this is the development of well-documented (Wang and Hwang 2001, Leckler et al. 2015) bimodal angular distributions of energy in highly peaked spectra as shown in the earlier work of Alves and Banner (2003).

**3.0 Development of Improved Theoretical Foundations for Source Terms**

Original arguments for transitioning to third-generation wave models were based on the need to represent all individual source terms with the same number of degrees of freedom as the directional spectrum being modeled (Hasselmann et al., 1985; WAMDIG, 1988). This premise implicitly assumes that representations of all three primary source terms are accurate in a detailed-balance context. As was shown in Figures 4 and 5, source terms in WW3 have been balanced to achieve accurate agreement with observed integrated parameters, even though individual source terms in the total balance can vary substantially while still producing reasonable agreement with such metrics based on such parameters. However, since the detailed frequency-direction structure of these terms exhibit significant variations, this is unlikely to be true for consistency with metrics based on spectral shape.

Given the increasing importance of wave models in coupled modeling systems and remote sensing, a number of groups within the NOPP program focused on the need to begin to develop improved detailed-balance source terms which use performance metrics based on spectral shape. A first step toward this goal was the development of a suitable set of metrics for such comparisons. Resio et al. (2016) summarized a number of studies to develop such a set of metrics based on a general consensus of measurements. (Toba, 1972; Donelan et al., 1985; Forristall, 1981; Ewans, 1998, Resio et al., 2004; Long and Resio, 2007; Romero and Melville, 2010).

Since several studies have shown that the current representation of nonlinear interactions is not in good agreement with the detailed-balance form of $S_{nl}$ (Resio and Perrie, 2008, Perrie and Resio, 2009; Resio et al., 2016), a major focus within the detailed-balance source term group has been investigations into three main areas:

1. Implementation of improved approximations to $S_{nl}$ into operational wave models

2. Testing new source terms with the full integral representation of $S_{nl}$

3. Development of improved theoretical foundations for source terms

**3.1 Implementation of improved approximations to $S_{nl}$ into operational wave models**

Considerable attention has now been focused on finding an improved representation of $S_{nl}$ in operational models. Two types of refinements have been investigated to date: 1) the Two-

Scale Approximation, or TSA, based on a decomposition of the spectrum into two scales of variability with a spectrum (Resio and Perrie, 2008; Perrie and Resio, 2009) and 2) extensions of the DIA to include additional discrete elements (Tolman, 2013). The TSA follows the structure of the WRT (Webb-Resio-Tracy) from Webb (1978) and Tracy and Resio (1982) integral form and does not include any tuning coefficients. Extensions of the DIA include the Multiple Discrete Interaction Approximation (MDIA, Hashimoto and Kawaguchi, 2002), which incorporates additional discrete interaction points that fall along the DIA's "figure-8" locus, and the Generalized Discrete Interaction Approximation (GMD) (Tolman (2013), which includes points that are not constrained to lie on the figure-8 locus (i.e. allow interactions among additional waves which meet the restrictions for resonant wave-wave interactions. Similar to the DIA, the MDIA and GDIA both rely on empirical tuning coefficients to optimize agreement to the full integral representation for a selected set of spectra. The MDIA and GDIA have been combined (and expanded upon) in the Generalized Multiple DIA (GMD) and exercised in holistic testing of performance relative to integrated parameters similar to those used in the previous section (Tolman, 2013). In the latter paper it is shown that a cascade of GMDs with increasing accuracy and costs could be created with objective tuning techniques for ST2. It is expected that similar cascades of GMDs with increasing complexity and cost can be derived for other source term packages, but there is no guarantee that the GMD results in universally tuned nonlinear interactions. A third approach, the Lumped Quadruplet Approximation (LQA) was initiated under this effort, but has not reached the stage of spectral testing and is not discussed here.

The TSA approach is based on a spectral decomposition into broad-scale (parameterized) and local-scale components and can be extended to multiple-peaked spectra (Perrie et al., 2013). It has been shown to provide significantly improved representation of $S_{nl}$ relative to the DIA in comparisons to the full integral. Perrie et al. (2013) adapted the TSA code to be executable within WW3 and tested its ability to reproduce idealized growth laws for dimensionless energy and peak frequency matching the WRT, using the ST4 wind input and wave dissipation source terms. Their work showed that the TSA produces essentially identical idealized growth curves in this comparison to the WRT, *and* that there are still notable differences in comparisons with one- and two-dimensional spectra, particularly for rapidly developing sea states. Perrie et al. (2013) also modified the TSA to include two separate broad-scale terms. This extension of the TSA, denoted dTSA, appears to improve the comparisons to measured significant wave heights off the coast of Nova Scotia during Hurricane Juan (Figure 6). However, it is important to realize that current wind input and wave dissipation source terms are optimized in conjunction with the DIA, which deviates significantly from both the WRT and the TSA; consequently, it is not surprising that changing only the nonlinear term alone will produce significantly improved results.

**3.2 Testing of wind input and wave breaking source terms with the full integral representation of $S_{nl}$**

Existing source terms in WW3 have been optimized to provide agreement with integrated parameters used to characterize wave heights, periods and directions. Several publications (Resio and Perrie, 2008; Perrie and Resio, 2009; Resio et al., 2016) have shown that the DIA and full integral representations of $S_{nl}$ differ markedly, with the DIA under-predicting $S_{nl}$ in young waves $c_p/u_{10} \ll 1$ and over-predicting $S_{nl}$ as waves approach full development $c_p/u_{10} \approx 1$. (Here, $c_p$ is the phase velocity at the peak frequency of the wave spectrum and $u_{10}$ is the 10-meter wind speed.) Rogers and van Vledder (2013) showed that the spectral widths generated by the same combination of $S_{ds}$ and $S_{in}$ using the full integral for $S_{nl}$ essentially removed the overprediction in the spectral widths in Lake Michigan given by the DIA-based model, but scatter was unfortunately worse, and so RMS-error was not improved. As noted previously, since existing representations of $S_{ds}$ and $S_{in}$ have been optimized to match integrated spectral parameters using the DIA, changing one term without re-balancing the entire source terms is not likely to produce much in the way of performance improvements

Banner and Morison (2010) continued to explore detailed-balance forms of $S_{in}$ and $S_{ds}$; however, much of their work shifted to the quantification of wave breaking probabilities as a function of the directionally-normalized spectral peak saturation, which provides an additional detailed-balance constraint. Further details are given in Banner et al. (2016), for which two journal papers are in preparation. Also, Banner and Morison (2018) revisit the upper ocean dissipation rate contribution from micro breakers and small whitecaps, which could be relevant to the high wavenumber cut-off in WW3.

## 3.3 Development of improved theoretical foundations

Wave generation is an extremely complex process, involving coupled motions between air and water which have very different densities. While excellent progress has been made in the prediction of integrated wave parameters, much of this progress has relied heavily on formulations containing multiple empirical coefficients to optimize the predictive capability for integrated spectral parameters within a specified basin. As was shown in the inter-comparison of different versions of WW3, markedly different combinations of source terms can reproduce somewhat similar performance for the significant wave height $H_s$ in basin-wide and global comparisons and in idealized tests, even when their detailed spectral shapes and source balances differ substantially. For this reason, it is essential to maintain firm theoretical grounds as the basis for developing future generations of models that are capable of accurately reproducing the characteristics of wave spectra found in nature. As part of the NOPP effort, two major efforts were conducted with this goal in mind: 1) a study by Pushkarev and Zakharov (2016) which investigated theoretical constraints on wind input source terms in fetch-limited situations and 2) a generalization of the kinetic equation (Hasselmann, 1962) to a generalized form that included additional spectral interactions neglected in the KE.

Pushkarev and Zakharov (2016) point out that the only source term used in spectral wave model derived directly from the Navier-Stokes equation with no empirical coefficients is the nonlinear wave-wave interaction term. They split the conventional integral for $S_{nl}$ into two parts,

$$S_{nl}(\omega_k,\theta_k) = F(\omega,\theta) - \Gamma(\omega,\theta)\varepsilon(\omega_k,\theta_k), \tag{3}$$

where the subscript "k" has been added to their notation to emphasize the point that the values $\omega$ and $\theta$ are identical in those two terms, whereas the functions $F(\omega,\theta)$ and $\Gamma(\omega,\theta)$ involve integrals over the full resonant interaction space. Since a stationary spectral form must have $S_{nl}(\omega,\theta) = 0,$ any solution of this equation must have the form,

$$\varepsilon(\omega_k,\theta_k) = \frac{F(\omega,\theta)}{\Gamma(\omega,\theta)} \quad \text{for } \Gamma(\omega,\theta) > 0 \tag{4}$$

consistent with what has become known as Kolmogorov-Zakharov solutions in many areas of physics.

Pushkarev and Zakharov also point out that for spectra in a fetch-limited situation, this adds a constraint on the sum of the three conventional source terms, such that

$$c_g \frac{\partial \varepsilon(\omega,\theta)}{\partial x} = S_{nl}(\omega,\theta) + S_{in}(\omega,\theta) + S_s(\omega,\theta) \tag{5}$$

where $c_g$ is the group velocity. The authors offer a good overview of the past work on the wind input term and perform tests to determine which of the many wind input terms available today are in reasonable agreement with empirical power laws for nondimensional wave parameters along a fetch. In these tests they found only Zakharov-Resio-Pushkarev (ZRP) wind input term satisfied all of the proposed test metrics.

A study of spectral shape characteristics showed that the DIA is unable to replicate the properties of shape evolution of spectral through time (Resio et al, 2016). Since part of the need to transition from second-generation to third-generation models was motivated by the need to have sufficient degree of freedom to allow a spectrum to evolve into a "natural" form (Hasselmann et al., 1985), this is a somewhat problematic result; but subsequent work by Ardag and Resio (in press) has shown that the assumption of an $f^{-5}$ (JONSWAP) form for its spectral basis, rather than an $f^{-4}$ form, results in this form providing fluxes that are inconsistent with those of the full integral.

The Generalized Kinetic Equation (GKE) was initially derived by Zakharov (1968). If one assumes that wave phases in the spectrum are negligibly correlated then the resonant (4-wave) interactions represent the lowest order interactions in the system. However, Annenkov and Shrira (2006) and Gramstad and Stiassnie (2013) have shown that at least two assumptions can be violated under certain circumstances: 1) the assumption that waves evolve on a slow time scale $O(\mu^{-4})$, where $\mu$ is wave steepness, and 2) the assumption that phases were uncorrelated over a substantial interval of time in the past.

In another set of idealized tests, Gramstad and Babanin (2016) performed a set of investigations focused on situations in which the KE might be expected to perform unsatisfactorily. As expected, one situation that the KE performs poorly compared to the GKE is for the case of unidirectional waves. Also, certain situations, such as an instantaneous change in wind direction showed substantial differences. However, overall the comparisons between the KE and GKE for realistic two-dimensional wave fields show good agreement with respect to the development of the main features of the spectral evolution.

**4.0 NOPP Shallow-Water Efforts**

In addition to predictions in open ocean areas, mission agencies supporting the NOPP effort identified a critical need for significantly improved wave forecasts in coastal and estuarine areas. These areas have three important differences that are typically not revealed in ocean-scale model comparisons:1) the importance of the frequency characteristics of wave breaking, which can significantly influence estimates of radiation stresses critical to surge predictions at the coast 2) wave generation in short, geometrically complex fetches (often in situations where the relative depth of the spectral peak is still considered quite deep and 3) wave generation and transformation on coastal shelves and nearshore areas, including estuarine areas where the depth becomes quite shallow and depth becomes an important factor in wave breaking. In general, compared with deep water wave modeling, shallow water modeling is complicated by increased importance of four factors, 1) the pronounced influence of spectral shape (particular directional characteristics) on wave conditions along the open coast; 2) complex short fetch geometries, which complicate the forms of source terms, 3) depth effects on propagation and source terms and 4) the effects of currents on wave propagations in localized areas, such as inlets.

Recent work identifies many potentially significant shallow water processes, such as: Bragg scattering by bottom irregularities (Ardhuin & Herbers, 2002), bottom friction, wave damping by viscous mud and vegetation (Ng, 2000, Dean and Bender, 2006) and eventually, when approaching the shoreline, triad wave-wave interactions (Freilich & Guza, 1984; Kaihatu & Kirby, 1995; Herbers & Burton, 1997; Agnon & Sheremet, 1997; Janssen et al. 2006; Smit & Janssen, 2016), and depth-induced breaking (Battjes & Janssen, 1978). In addition to these physical processes, there are also many practical issues that complicate wave modeling in shallow water. Specifically: the need for accurate bottom bathymetry data, bottom roughness information, tidal elevations and currents, and adequate prediction of tides, storm surges, seiches, and wave- and wind-driven set-up during storms.

Efforts to include parameterizations for the dissipative processes (e.g. friction, wave breaking and interaction viscous mud or vegetation) in the RTE have been relatively straightforward because the resulting approximations take the form of source terms that are readily included into the RTE framework. Nevertheless, our understanding of these processes (in particular wave breaking) is still very incomplete, and improvements in the parameterizations need to be actively pursued.

Accounting for wave inhomogeneity or triad non-linear interactions within the RTE paradigm (Eq. 1, 2), can be very problematic in some circumstances because the RTE is based on the premise that the wave field is quasi-homogeneous and near-Gaussian. This implies that each spectral component is only weakly coupled to other components and can be treated under the assumption that higher order moments of the surface heights and velocities can be neglected. However, this is not always the case, for example focusing, shoaling and other shallow-water processes can lead to statistical inhomogeneities through correlations among non-collinear waves. Further, the amplification of lower-order (three-wave) nonlinearity in shallow water, can lead to the deformation of the wave shape, as well as the generation of infragravity wave motion (e.g., Elgar et al., 1992, 1994; Herbers et al., 1994, 1995a,b; Sheremet et al., 2002, 2005; Henderson et al., 2006; Janssen et al., 2003, 2006; Ruessink, 1998, and many others) and a departure from Gaussian statistics of water elevations around the mean. The evaluation of such wave inhomogeneity requires cross-correlations to be estimated (Elgar and Guza, 1985; Kaihatu and Kirby, 1995; Kaihatu and Tahvildari, 2012; Smit & Janssen, 2016). Since the conventional RTE model only considers the transport of the variance-carrying components of the second-order correlation matrix, these effects (inhomogeneity and non-Gaussianity) are neglected.

Fundamental improvements to include higher-order coastal wave dynamics in stochastic models requires a generalization of the RTE framework to propagate cross-correlators, where necessary (potentially affecting the left side of Equation 1) and additional transport equations for the associated higher-order correlations. In this project, a fundamental re-derivation of the RTE was numerically implemented in a research stochastic model to validate the potential of these approaches. To terminate the expansion of terms in the high-order moments in spectral models, requires a closure argument which introduces additional assumptions related to this approximation. Alternatively, such nonlinear effects can be accurately modeled with phase-resolved models (using ensemble averaging across multiple realizations). Under a range of propagating wave spectra, such models can accurately model near-resonant nonlinear effects due to triad interactions, and can provide a benchmark solution for validation and calibration of nonlinear and complex wave-bottom interaction processes in stochastic models. In very shallow water, and very near the coast, where spatial scales are typically smaller, deterministic or phase-resolved models are a viable alternative and can help improve stochastic models (see e.g. Agnon & Sheremet, 1997; Herbers & Burton, 1997; Janssen et al. 2006, Davis et al., 2014; Sheremet et al., 2016).

In this project, both stochastic and phase-resolved approaches were explored to advance modeling of shallow-water waves, and enable a seamless transition from deep to shallow water.

To explore the potential of deterministic phase-resolving models, an implementation of the angular spectrum model proposed by Agnon & Sheremet (1997) is applied to investigate the effects of nonlinear wave propagation, visco-elastic materials on the seafloor, and interactions with vegetation. The implementation of phase-resolved approximations were not expected to yield operational models in the time frame of the project but were intended only to provide a benchmark for complex dissipation and nonlinear processes.

Within the NOPP framework, stochastic developments focused on developing generalizations of the RTE framework that could be integrated into existing stochastic models to provide an immediate extension of operational models to connect deep-to-shallow water. This led to the development of a generalized form of the RTE which fully accounts for the transport equations for wave cross-correlations (see Smit & Janssen, 2013, 2016; Smit et al. 2015). This fundamentally new result based on first principles enables RTE-type models to transport second- and third-order cross-correlators to account for inhomogeneity and nonlinear effects in coastal waters. Further, improved source terms for shallow-water wave breaking were developed and tested extensively across a wide range of conditions (e.g. Salmon et al. 2015), improvements to triad formulations have been developed and tested (Salmon et al. 2016), and an efficient quadruplet formulation has been developed and tested.

Since both approaches are fundamentally different (deterministic versus stochastic), we subdivide the following into a section on Stochastic Modeling Advances, and Phase-Resolved Modeling Advances to outline the main results of the work in this project. For details the reader is referred to the cited publications.

**4.1 Development of Improved Stochastic Modeling Capabilities**

**4.1.1 Stochastic modeling of inhomogeneous wave fields**
The modeling of spatially inhomogeneous wave fields in shallow water requires the development and transport of wave-cross-correlators between waves that propagate in different directions. Because there are no conservation principles available for such cross-correlations, transport equations were derived staring from the underlying equations of motion (Smit & Janssen, 2013). Through a series of operations, and applying the usual WKB approximation to incorporate the slowly varying medium, the resulting transport equation for the wave statistics, which includes coherent interference effects (including diffraction), can be written in the form (see Smit et al. 2015a)

$$\partial_t \mathcal{E} + \mathbf{c_x} \cdot \nabla_\mathbf{x} \mathcal{E} + \mathbf{c_k} \cdot \nabla_\mathbf{k} \mathcal{E} = S_{SC} + S_{DS} \qquad (6)$$

where the transported variable $\mathcal{E}$ is referred to as the Coupled-Mode spectrum (Janssen et al. 2008). On the right side of equation 6, the scattering term $S_{SC}$ and dissipative source term $S_{DS}$ allow for the development of coherent interferences through interaction with the variable medium and wave breaking dissipation in a coherent wave field. In general, these terms take the form of a convolution integral of the form (Smit et al. 2015a,b)

$$S_j = \int L_j(\mathbf{k},\mathbf{q},\mathbf{x},-i\nabla_\mathbf{x})\mathcal{E}(\mathbf{k}-\mathbf{q},\mathbf{x})d\mathbf{q} \qquad (7)$$

where, for $S_{SC}$ and $S_{DS}$, $L_j$ represents a either a dispersion or dissipation forcing term, respectively [see Smit et al. 2015a,b for details]. Equation 7, which we refer to as a Quasi-Coherent (QC) model, closely resembles the standard RTE (eq. 1). However, there are also several fundamental and important differences. Principally, the Coupled-Mode (CM) spectrum $\mathcal{E}(\mathbf{k},\mathbf{x},t)$ is not a variance density spectrum. Instead it represents the complete second-order statistics, including both variance-carrying components of the spectrum (the variance density spectrum), and cross-correlations. As a consequence, the transport equation 6 is in essence a generalization of the RTE, and transports a more complete spectral representation of the wave statistics in cases where higher order statistics are important. Through the inclusion of cross-correlations, this approximation has the potential to improve estimates of conventional bulk wave statistics (e.g. significant wave height) in regions of strong wave interference (Figure 7), while reducing to the RTE in regions where the wave field is practically homogeneous.

Since equations 6 and 7 are a natural extension of RTE-type model equations, these advances can be seamlessly integrated into operational models once validated in operational applications, and can be applied at typical coastal scales (Smit et al. 2015a, Smit & Janssen, 2016). It is thus a natural extension for operational models to bridge the gap from deep-to-intermediate water. These new equations have been tested on regional scales (Smit et al. 2015a) and development on a full integration into SWAN are ongoing.

### 4.1.2 Stochastic modeling of non-Gaussian wave fields

To model non-Gaussian statistics associated with non-linear interactions, a stochastic model requires a transport equation for the three-wave correlator, or bi-spectrum. However, since the bi-spectrum is essentially a higher-order cross-correlator, the approach developed to transport cross-correlations for inhomogeneous (but Gaussian) wave fields (Smit & Janssen, 2013), can likewise be applied. To that end we start out with a nonlinear version of the equation of motion (Janssen et al. 2006), and with the same method used to derive the QC model for inhomogeneous wave fields, we derive – assuming the wave field to be non-Gaussian but homogeneous - the coupled set of equations for the spectrum and bispectrum, which can be written as (Smit & Janssen, 2016)

$$\partial_t \mathcal{E} + \mathbf{c}_\mathbf{x} \cdot \nabla_\mathbf{x} \mathcal{E} + \mathbf{c}_\mathbf{k} \cdot \nabla_\mathbf{k} \mathcal{E} = S_{NL3}(\mathcal{B}) \qquad (8)$$

$$\left[\partial_t + \mathbf{c}_\mathbf{x}^{\mathbf{k},\mathbf{k}'} \cdot \nabla_\mathbf{x} + \mathbf{c}_\mathbf{k}^{\mathbf{k},\mathbf{k}'} \cdot \nabla_\mathbf{k} + \mathbf{c}_{\mathbf{k}'}^{\mathbf{k},\mathbf{k}'} \cdot \nabla_{\mathbf{k}'}\right]\mathcal{B} = R_{\mathbf{k},\mathbf{k}'}\mathcal{B} - iQ(\mathcal{E}) + iC \quad . \qquad (9)$$

Here $S_{NL3}$ is the triad wave-wave forcing term, which depends on the local bispectrum $\mathcal{B}$. Further, $\mathbf{c}_\mathbf{x}^{\mathbf{k},\mathbf{k}'}$, $\mathbf{c}_\mathbf{k}^{\mathbf{k},\mathbf{k}'}$, and $\mathbf{c}_{\mathbf{k}'}^{\mathbf{k},\mathbf{k}'}$ are transport velocities through geographical and wavenumber space respectively [see Smit & Janssen, 2016 for their definition], $R$ is a linear term to account for resonance mismatch and shoaling effects, the $Q_{k,k'}$ contains products of spectrum components, and $C$ is the fourth-cumulant. The transport equation for the bispectrum, equation 8, is referred to as the bi-radiative transport equation (bRTE) (Janssen et al., 2008). The bRTE allows the fully two-dimensional evolution of the bi-spectrum through a slowly varying medium, while being consistent with the assumptions underlying the RTE, and without introducing additional constraints on bandwidth, aperture, or medium variability (Smit & Janssen, 2016). As such, the bRTE can be readily coupled to a conventional stochastic model based on the RTE, such as SWAN or WW3.

## 4.2 Development of Improved Phase-Resolved Modeling Capabilities

### 4.2.1 Propagation Effects
A Monte Carlo approach to nonlinear shoaling evolution of waves using a unidirectional version of the TRIADS model (Davis et al., 2014; Sheremet et al., 2016) was developed within the NOPP effort. This allows direct integration of the directional, hyperbolic formulation of the phase-resolving triad interaction proposed by Agnon and Sheremet (1997). The formulation assumes the beach to be laterally uniform and mildly sloping in the cross-shore direction. The model accounts for refraction, shoaling, and nonlinear triad interactions in shoreward propagating waves. While, in general, near-resonant triads dominate the evolution in certain situations over the short spatial scales of a typical beach (on the order of 10 wavelengths, e.g. Agnon and Sheremet, 1997), 4-wave interactions are still expected to dominate the dynamics of high-frequency modes, and play an important role in the nonlinear shoaling of coherent structures. Parametrizations of wave dissipation/growth processes can be readily incorporated.

An example of a simulation of a directional spectrum from Hurricane Bill at the US Army Corps field measurement site at Duck, North Carolina is provided in Figure 8. The simulations compare reasonably with the observations both in directional character and frequency distribution, capturing the overall shape of the spectrum, and also describe more subtle nonlinear trends such as the growth of the Infra Gravity (IG) band cross-shore energy flux, the broadening of the frequency spectrum, and the generation of peak harmonics. Duck, NC is particularly suited to use this model because it is very close to a gently sloping one-dimensional profile. This effort shows that more effort is needed to combine multiple nonlinear transfer mechanisms to obtain realistic wave spectra near the coast. Unlike 4-wave interactions, 3-wave interactions can distort the gaussian distribution of near-bottom orbital motions which can influence sediment transport in such areas.

## 4.2.2 Dissipation due to Bottom Friction on Sandy Coasts, Mud, and by Vegetation

Shallow coastal and shelf areas are found distributed around the globe. Besides the dissipative effects of energy loss mechanisms in shallow water, many of the inland areas adjacent to these areas are also protected by widespread vegetation (marsh, sawgrass, mangroves, etc) during high water – high wave events.  In the US, the areas of the Louisiana coast are significantly influenced by the natural protection provided by both of these wave-height reduction mechanisms.  The Gulf Coast Team sought to extend the TRIADS models to these regions via addition of terms to account for dissipation due to vegetation. Wave damping by mud affects both the dynamics of the waves (by reducing wave energy) and the kinematics (by affecting the wave number).  The former of these effects is well known, but the latter is little explored. This becomes an issue with obliquely- approaching waves, where refraction is an important component of the wave field; the effect of bottom mud on these processes is unclear.

Wave-current interaction is a persistent feature in many estuarine environments with water levels modulated by tides. These areas are marked by vast expanses of bottom mud, making the study of mud-induced dissipation combined with wave-current interaction very important to predictions of waves and currents. Kaihatu and Tahvildari (2012) used a one-dimensional nonlinear shoaling model (Kaihatu and Kirby, 1995), extended to include wave-current interaction (Kaihatu, 2009) and viscous mud-induced dissipation (Ng, 2000). Data from An and Shibayama (1995) and Zhao et al. (2006) were used to validate the model. Both experimental studies assumed the mud acted as a viscoelastic material, so the viscous dissipation mechanism of Ng (2000) was calibrated with the data to match the measured dissipation rate. Additionally, the resulting model was used to investigate the effect of following and adverse currents on mud-induced wave dissipation. It was found that following currents (currents in the same direction as wave propagation) tended to reduce the rate of mud-induced dissipation, while adverse currents tended to enhance the dissipation, consistent with observations by An and Shibayama (1995) and Zhao et al. (2006). Figure 9 shows wave spectral evolution over a flat bottom and a highly dissipative mud for different Ursell numbers and Froude numbers. It is apparent that adverse wave-current interaction enhances wave dissipation by mud, while following currents mitigate dissipation. Further testing of this basic process model is needed, but when complete can be incorporated into both TRIADS for further testing and parameterization for phase-averaged predictive models.

The influence of vegetation in estuarine environments was also investigated as part of this overall study. Dean and Bender (2006) assumed that dissipation due to drag on individual stem elements can be parameterized as a function of vegetative stem diameter and associated stem spacing. This mechanism was incorporated into the parabolic model of Kaihatu and Kirby (1995). Venkattaramanan (2014) used the dissipation mechanism of Kobayashi et al. (1993) to represent vegetation dissipation in the one-dimensional model of Kaihatu and Kirby (1995). This mechanism assumed an exponential decay of wave energy due to rigid vegetation elements, and is expressed in terms of a Reynolds-number-dependent drag coefficient. To offset the limitation to rigid elements (which greatly limits the applicability of the model) the drag formulation of Mendez et al. (1999) was used for drag coefficient specification, which allows for some degree of vegetation motion. Figure 10 shows a comparison between the model and laboratory data of Anderson and Smith (2013). Wave spectra compare well except for some aspects of the high frequencies, while the root-mean-square wave heights are a very good match to data. As with the wave-current-mud interaction, vegetation effects can be incorporated into the TRIADS model for further testing and parameterization for later inclusion into phase-averaged wave prediction models.

Routines to estimate dissipation of waves by a viscous mud layer was implemented in the phase-averaged models SWAN and WW3, and dissipation of waves by vegetation was implemented in SWAN. The Ng (2000) mud dissipation formulation was implemented in SWAN (Rogers and Holland 2009, official release 41.01 in 2014) and in WW3[3] (Rogers and Orzech 2013, official release 4.18 in 2014). The vegetation dissipation formulation of Dalrymple et al (1984) was added to SWAN (Suzuki et al 2012, official release 40.81 in 2010). The difficulty of objective prescription of required inputs remains a major obstacle to operational use of these source terms.

## 5.0 Discussion: Project Metrics Accomplished

Goals of theproject met through the research program include the following:

- Two Source Term Sets (ST4 and ST6) fast enough for operational forecasting were developed and tested. A third (ST5) underwent more limited testing due to its use of the full WRT $S_{nl}$ algorithm.
- In global evaluations the ST4 and ST6 were shown to reduce wave height errors by 30-50%, primarily due to improvements in deep-water swell heights compared to earlier versions of WW3.

---

[3] The WW3 implementation also included the dissipation formulation of Dalrymple and Liu (1978) as an additional option

- ST4 and ST6 were incorporated into the WW3 and SWAN models and made available to these user groups.
- A new nonlinear source term based on the Generalized Kinetic Equation was introduced and tested against the WRT algorithm.
- Investigations into wave spectral characteristics demonstrated that $S_{nl}$ in present model do not show consistency with observed spectral shapes.
- The National Weather Service and the Navy's Fleet Numerical Meteorology and Oceanography Center made ST4 the operational version of theWW3.
- Additional improvements were made to the breaking and triad source terms of the SWAN model as well as for coupled wave-bottom interactions.
- Additional improvements were made to theWW3 model for swell dissipation and breaking.
- A new stochastic framework was developed to transport cross-correlations in RTE-type models, which enable modeling of wave inhomogeneity and nonlinearity in such models, and thus provide a seamless transition from deep-to shallow water evolution in operational models.
- A phase-resolved model (TRIADS) is implemented to study physics of nonlinear wave evolution in shallow water, and allow accurate prediction of sub- and superharmonics. The TRIADs model has been extended to treat muddy environments and propagation through vegetation.

The decision to use theWW3 and SWAN systems as integrated development platforms provided an efficient mechanism for exchange of code and to perform evaluations. As a result NWS and FNMOC were able to adopt ST4 for operational use before the end of the program. The rapid transition of the algorithms to theWW3 and SWAN user groups comprising together some 3000 users world-wide provided the wave research and engineering communities with robustly evaluated state of the art technology. The researchers and users have the option of using one of the new versions ofWW3 or SWAN, or of extracting specific algorithms.

Throughout the paper suggestions have been made concerning pathways for future research. Here we will summarize the more general conclusions.

The significant improvement in the Global Forecast statistics in ST4 and ST6 come from better estimation of swell dissipation and by bounding the input and output momentum and energy to be consistent with drag laws and breaking estimates. So in an integral sense the model captures the net energy change. The idealized growth tests show that although the growth of wave height may be similar in the models, the details of where the individual source terms add and subtract energy markedly differ excepting the nonlinear source term, the DIA, which was used in both sets of source terms. This inconsistency suggests that the underlying physics of the source terms for input and dissipation are still not understood.

The comparisons of Stopa et al. (2016) indicated poorer performance on prediction of directional properties when coupled with the differences in source term formulations suggests that future evaluations should be directed at spectral comparisons. The central tenet of third generation wave modeling such as WW3, WAM and SWAN is the requirement to compute the directional spectrum. Thus work is needed on a better representation of the physics in this dimension as well as developing data sets of highly resolved directional spectra.

For shallow and coastal areas emphasis was placed on fundamentally extending RTE-based models into shallow water to the beach to allow modeling of inhomogeneity and non-Gaussian effects in operational models. In addition, a phase-resolved model was implemented to study nonlinear dynamics and strong interaction when dissipative mechanisms (mud, vegetation) are dominant.

The extensions developed to the RTE framework in this project enable modeling of inhomogeneous effects (focusing, refraction, diffraction) and nonlinearity in stochastic models. This is a fundamental departure of classic RTE-type models. The stochastic models developed here include transport of cross-correlations, which fully account for effects such as diffraction, and the development non-Gaussian statistics, while still being fully compatible with RTE-type models. After initial testing of these formulations on regional scales, they are now being fully implemented into SWAN to create the next generation shallow-water stochastic model. This work is ongoing and supported by NWO, the Netherlands (project ALWOP.167).

In the deep-water discussion a major improvement was in part ascribed to constraining the model's input by a drag law and by tying dissipation to the form of the spectral tail. In shallow water the wave field is often dominated by swell, which is generally thought to modify drag. Likewise dissipation in shallow water becomes much influenced by bottom type. So compared to the deep water situation, the constraints for shallower water may be poorly known and perhaps be more site and wave specific than the deep-water case.

Our overview suggests however that for coasts and shores that have more heterogeneous depth fields and bottom material distributions the problem is very complex. Lack of knowledge of these factors may degrade even a well-formulated model prediction. In the earlier section we discussed the need to have comprehensive measurements to evaluate a model and suggested that an observing approach including remote and in situ data sets, coupled with data assimilation and inverse modeling capabilities may be an effective way to develop site specific models data sets. Use of Monte Carlo techniques varying the unknown factors in a consistent manner may also provide a probabilistic prognostication.

In the end, the approaches in deep and shallow water rely on an ability to predict how the energy is distributed in direction and by frequency (or wave number). In shallow water in particular accurate prediction of the wave field can be sensitive to correct specification of the directional spectrum of the inbound wave field. This then provides an important reason for research on the deep water modeling focusing on prediction of the spectrum, not just integral parameters such as wave height.

**6.0 Summary and Future Directions**

This paper provides an overview of the efforts of investigators funded under the National Oceanographic Partnership Programs Wave Model research program. The program produced new sets of balanced source terms that were evaluated with extensive data sets and which reduced wave model errors compared to previous versions of these codes. NOAA and FNMOC adopted one of the source terms for their operational version of WW3. The source terms were added to the WW3 and SWAN models, making these improvements available to scientists worldwide.

In Shallow Water and Coastal environments, a number of theoretical and numerical upgrades have been introduced; however, none of these have yet been validated for operation applications. Among these developments are the following:

1. Upgrades to SWAN were made to improve breaking and the generation of near-resonant harmonics.

2. A unified shallow to deep water breaking algorithm was added toWW3.

3. A new stochastic modeling framework was developed to transport cross-correlators and enable modeling of inhomogeneity, diffraction, and non-Gaussian statistical characteristics in stochastic models from first principles, potential extending the RTE framework to include higher order moments in wave characteristics.

4. The TRIADS model was used to investigate nonlinear shoaling and dissipation over mud and vegetation using Monte Carlo approaches to account for wave phase variations.

As with most papers addressing progress in a particular field, the focus here has been on positive aspects of accomplishments. This naturally addresses aspects of improvements in wave modeling and does not dwell on the challenges that remain. The original motivation for the need to transition from second-generation models to third-generation models was to ensure that the physics would allow spectra to relax to their natural form in order to achieve a universal form for wave generation, dissipation and propagation and, in contrast to 2G models, to obtain a universal form for relevant physical processes that did not need site specific retuning. In the 30 years that have passed since 3G models were first implemented for operational applications, many papers have introduced variations in source terms and have shown improvements in comparisons between predicted and measured wave parameters such as significant wave height for sea and swell, mean wave period, peak wave period, mean direction and directional spreading; however, they have not focused on developing an understanding of potentially critical issues that remain.

Investigations into problem areas are often difficult inside a scientific community, since they can create internal stresses among different groups and can challenge fundamental assumptions within operational models. Because of this, it is essential that such an undertaking be organized with a well-defined purpose at its initiation. Careful attention to the metrics for such model evaluations is critical. It is clear from results presented here that operational models provide a reasonable approximation to most day-to-day wave conditions in oceanic areas. However, many of the applications of wave models today are used to predict conditions during extreme events and in areas which may not be directly connected to oceans. This context brings us back to the reason for the initial development of 3G models - the need for understanding the fundamental physics of arbitrary-depth, detailed-balance form.

Suggestions for such a collaborative effort include the following:

1. Investigation of model performance in the prediction of wave spectra during extreme events: Model performance during such events represent a miniscule portion of the comparison data and do not have much, if any, influence on operational metrics used in most oceanic-scale model evaluations, such as those presented here. Since many of the critical applications of wave models involve their use for planning and design this is a clear deficiency in ongoing operational testing.

2. Improved physics in source terms: Many of the existing models now contain empirical coefficients, which are tuned differently for different basins. The ability for models to function consistently over a range of scales, without varying empirical coefficients, was the initial goal of third-generation model, since there should be no need to retune models for applications in different basins if the physics is properly posed.

3. Investigation of modeled spectral characteristics: Although some operational applications can be satisfied with metrics based only of integrated wave parameters of wave height period and direction in deep water, the distribution of energy in frequency and direction is critical as boundary conditions for nearshore wave and surge predictions. It is also very important to meet the needs of applications in coupled modeling systems,

3. Investigation into details of the transformation in shallow water and its inclusion into operational models: The schism between deep-water and shallow-water regimes in this paper needs to be rectified in order to help meet the needs of coastal communities in rising sea level and changing climate. A systematic, objective comparisons of shallow-water wave predictions should be undertaken with the same level of effort and support as has been devoted to deep-water model performance.

# REFERENCES

Agnon, Y., and Sheremet, A., 1997: Stochastic nonlinear shoaling of directional spectra. *J. Fluid. Mech.* **345**, 29-58.

Aijaz, S., Rogers, W.E., and A. Babanin, 2016: Wave spectral response to sudden changes in wind direction in finite-depth waters. *Ocean Mod*, **103**, pp 98-117.

Alves, J.H., and Banner, M.L. 2003: Performance of a saturation-based dissipation rate source term in modeling the fetch-limited evolution of wind waves. *J. Phys. Oceanogr.* **33**, 1274-1298.

An, N., and Shibayama, T., 1995: Wave-Current Interaction with Mud Bed. *Coastal Engineering*. pp. 2913-2927. doi: 10.1061/9780784400890.211.

Anderson, M.E., and Smith, J.M., 2013: Wave attenuation by flexible, idealized salt marsh vegetation. *Coast. Engrg.* **83,** 82-92.

Andrews, D. G., and M. E. McIntyre, 1978: On wave-action and its relatives. *J. Fluid Mech.*, **89**, 647–664, doi:https://doi.org/10.1017/S0022112078002785.

Annenkov, S. Yu. & Shrira, V. I., 2006: Direct numerical simulation of downshift and inverse cascade for water wave turbulence. *Phys. Rev. Letters*, **96**(20), 204501(4).

Ardhuin, F. and T. H. C. Herbers, 2002: Bragg scattering of random surface gravity waves by irregular sea bed topography. *J. Fluid Mech.*, **451**, 1–33.

Ardhuin, F., T.H.C. Herbers, G.Ph. van Vledder, K.P. Watts, R. Jensen, and H.C. Graber, 2007: Swell and slanting fetch effects on wind wave growth. *J. Phys. Oceanogr.*, **37**(4), pp 908-931.

Ardhuin, F., L. Marié, N. Rascle, P. Forget, and A. Roland, 2009b: Observation and estimation of Lagrangian, Stokes, and Eulerian currents induced by wind and waves at the sea surface. *J. Phys. Oceanogr.*, **39,** 2820–2838.

Ardhuin F., and Coauthors, 2010: Semiempirical Dissipation Source Functions for Ocean Waves. Part I: Definition, Calibration, and Validation. *J. Phys. Oceanogr.*, **40**, 1917–1941, doi:10.1175/2010JPO4324.1.

Ardhuin, F., Tournadre, J., Queffelou, P., and F. Girard-Ardhuin, 2011: Observation and parameterization of small icebergs: drifting breakwaters in the southern ocean

*Ocean Modell.,* **39** (2011), pp. 405-410.

Ardhuin, F., Dumas, F., Bennis, A.C., Roland, A., Sentchev, A., Forget, P., Wolf, J., Girard, F., Osuna, P., and Benoit, M., 2012: Numerical wave modeling in conditions with strong currents: dissipation, refraction and relative wind. *J. Phys. Oceanogr.*, **42**, 2101–2120.

Babanin, A. V., K. Tsagareli, I. Young, and D. Walker, 2007: Implementation of new experimental input/dissipation terms for modelling spectral evolution of wind waves. *Proc. 10th Int. Workshop on Wave Hindcasting and Forecasting,* Oahu, HI, WMO/IOC Joint Technical Commission for Oceanography and Marine Meteorology (JCOMM), C2. [Available online at http://www.waveworkshop.org/10thWaves/ProgramFrameset.htm].

Babanin, A.V., Chalikov, D., Young, I. and I. Savelyev, 2010: Numerical and laboratory investigation of breaking of steep two-dimensional waves in deep water. *J. Fluid Mech.*, **644,** 433–463.

Babanin, A.V., 2011: Breaking and dissipation of ocean surface waves. *Cambridge University Press,* p 480.

Babanin, A.V., 2012: Swell dissipation function for wave models. *Geophysical Research Abstract: 2012* European Geosciences Union General Assembly (EGU), Vienna, Austria, 22-27 April 2012, Vol **14**.

Banner, M. L., and I. R. Young, 1994: Modeling spectral dissipation in the evolution of wind waves. Part I: Assessment of existing model performance. *J. Phys. Oceanogr.*, **24,** 1550–1570.

Banner, M. L., Babanin, A. V., Young, I. R., 2000: Breaking probability for dominant waves on the sea surface. *J. Phys. Oceanogr*. **30**, 3145–3160.

Banner, M.L. and R.P. Morison, 2010: Refined source terms in wind wave models with explicit wave breaking prediction. Part I: Model framework and validation against field data. *Ocean Modell.* **33** (1-2), 177-189. http://dx.doi.org/10.1016/j.ocemod.2010.01.002.

Banner, M. L., Barthelemy, X., Fedele, F., Allis, M., Benetazzo, A., Dias, F., and W. L. Peirson, 2014a: Linking reduced breaking crest speeds to unsteady nonlinear water wave group behavior. *Phys. Rev. Lett.*, **112**, 114502, doi: https://doi.org/10.1103/PhysRevLett.112.114502

Banner, M. L., Zappa, C.J. and J. Gemmrich, 2014b: A note on Phillips' spectral framework for ocean whitecaps. *J. Phys. Oceanogr.*, **44**, 1727–1734, doi:https://doi.org/10.1175/JPO-D-13-0126.1.

Banner, M.L., R.P. Morison and C.W. Fairall, 2016: Refined Source Terms in Wave Watch III with Wave Breaking and Sea Spray Forecasts. Final report to NOPP DRI: Improving Wind Wave Predictions: Global to Regional Scales. http://www.dtic.mil/dtic/tr/fulltext/u2/1014558.pdf

Banner M.L. and Morison R.P., 2018: On the upper ocean turbulent dissipation rate due to microscale breakers and small whitecaps, *Ocean Modelling*, vol. **126**, pp. 63 - 6, http://dx.doi.org/10.1016/j.ocemod.2018.04.004.

Barnett, T.P., 1968: On the generation dissipation, and prediction of wind waves. *J. Geophys. Res.* **73**, 529-531.

Battjes, J.A., and J.P.F.M. Janssen, 1978: Energy Loss and Set-up Due to Breaking of Random Waves, Proc. 16th Int. Conf. Coastal Engineering, *ASCE*, pp. 569-587.

Bidlot, J., Janssen, P., Abdalla, S., 2007: A revised formulation of ocean wave dissipation and its model impact. Tech. Rep. Memorandum 509, ECMWF, Reading, U. K.

Bidlot, J.R., 2012: Present status of wave forecasting at E.C.M.W.F. *Proc. ECMWF Workshop on Ocean Waves*, Reading, United Kingdom, ECMWF, 16 pp. [Available online at http://old.ecmwf.int/publications/library/ecpublications/_pdf/workshop/2012/Ocean_Waves/Bidlot.pdf.]

Booij, N, Ris, R.C., and L.H. Holthuisen, 1999: A third-generation wave model for coastal regions, Part I, Model description and validation. *J. Geophys. Res*, **104**, 7649 – 7666.

Caires, S. & Sterl, A., 2005: A New Nonparametric Method to Correct Model Data: Application to Significant Wave Height from the ERA40 Re-Analysis. Journal of Atmospheric and Oceanic Technology - *J Atmos Ocean Technol*. **22,** 443-459. 10.1175/JTECH1707.1.

Chawla, A., Wilson-Diaz, D. & Tolman, H., 2013: Validation of a thirty year wave hindcast using the Climate Forecast System Reanalysis winds. *Ocean Modelling*. **70**, 189-206. 10.1016/j.ocemod.2012.07.005.

Chen, G., Chapron, B., Ezraty, R. and D. Vandemark, 2002: A global view of swell and wind sea climate in the ocean by satellite altimeter and scatterometer. *Am Meteorol. Soc.,* November 2002, pp 1849-1859.

Dalrymple, R.A., Kirby, J.T., Hwang, P.A., 1984: Wave diffraction due to areas of energy dissipation. *J. of Waterway, Port, Coastal and Ocean Eng*., **110**, Issue 1.

Dalrymple, R.A. and P.L.F. Liu, 1978: Waves over soft muds: A two-layer fluid model. *Am. Meteor. Soc*, **8**, 1121-1131.

Davis, J.R., Sheremet, A., Tian, M., and Saxena, S., 2014: A numerical implementation of a nonlinear mild slope model for shoaling directional waves. *J. Marine Science and Eng*. **2,** 140-158.

Dean, R.G., and C.J. Bender, 2006: Static wave setup with emphasis on damping effects by vegetation and bottom friction. *Coastal Engineering,* **53**, (2–3), 149–156.

Donelan M. A., J. Hamilton, and W. H. Hui, 1985: Directional Spectra of Wind-Generated Waves. *Philos. Trans. R. Soc. A Math. Phys. Eng. Sci.*, **315**, 509–562, doi:10.1098/rsta.1985.0054.

Donelan, M. A., 1999: Wind-induced growth and attenuation of laboratory waves. In: Sajjadi, S. G., H.Thomas, N., Hunt, J. C. R. (Eds.), Wind-overwave couplings. Clarendon Press, Oxford, U. K., pp. 183–194.

Donelan, M.A., Babanin, A.V., Young, I.R., Banner, M.L., McCormick, C., 2005: Wave-follower field measurements of the wind-inputspectral function. Part I: measurements and calibrations. *J. Atmos. Ocean. Technol.,* **22,** 1672-1689.

Donelan, M.A., Babanin, A.V., Young, I.R., Banner, M.L., 2006: Wave-follower field measurements of the wind-input spectral function. Part II: Parameterization of the wind input. *J. Phys. Oceanogr.,* **36,** 1672-1689.

Edwards, K.L., W.E. Rogers, S. Siqueria, P. Gay, K. Wood, 2018: A cost-benefit analysis of SWAN with source term package ST6. NRL Memorandum Report, NRL/MR/7320-18-9755, 41pp. (http://www7320.nrlssc.navy.mil/pubs.php)

Elgar, S., Herbers, T.H.C., Okihiro, M., Oltman-Shay, J., and R.T. Guza, 1992: Observations of infragravity waves. *J. Geophys. Res.,* **97**, 15, 573-577.

Elgar, S., Herbers, T.H.C., and R.T. Guza, 1994: Reflection of ocean surface gravity waves from a natural beach. *Am. Meteorological* Soc.*, 1503-1511.*

Elgar, S., Guza, R.T., 1985: Observations of bispectra of shoaling surface gravity waves, *J. Fluid Mech.,* **161**, 425–448*.*

Ewans, K. C., 1998: Observations of the Directional Spectrum of Fetch-Limited Waves. *J. Phys. Oceanogr.*, **28**, 495–512, doi:10.1175/1520-0485(1998)028<0495:OOTDSO>2.0.CO;2.

Ewing, J.A., 1971: A numerical wave prediction method for the North Atlantic Ocean. *Dtsch. Hydrogr.* Z 24, 241-262.

Fairall, C.W., Banner M.L., Peirson W.L., Asher, W., and Morison R.P., 2009: Investigation of the physical scaling of sea spray spume droplet production, *Journal of Geophysical Research: Oceans*, vol. **114**

Filipot, J.F., Ardhuin, F., 2012: A unifi.ed spectral parameterization for wave breaking: from the deep ocean to the surf zone. *J. Geophys. Res*., **117**, C00J08.

Forristall, G. Z., 1981: Measurements of a saturation range in ocean wave spectra. *J. Geophys. Res.*, **86**, 8075-8084.

Frielich, M.H., Guza, R.T., 1984: Nonlinear effects on shoaling surface gravity waves. *Royal Society Publishing,* **311**, 1-41.

Gramstad, O., Stiassnie, M., 2013: Phase-averaged equation for water waves. *J Fluid Mech.,* **718**:280–303.

Gramstad, O., & Babanin, A., 2016: The generalized kinetic equation as a model for the nonlinear transfer in third-generation wave models. *Ocean Dynamics*, **66**: 509. doi:10.1007/s10236-016-0940-4.

Hashimoto, N. & K. Kamaguchi, 2002: Extension and modification of discrete interaction approximation (DIA) for computing nonlinear energy transfer of gravity wave spectra. Presented at the 4[th] International Symposium on Ocean Wave Measurement and Analysis. https://doi.org/10.1061/40604(273)55.

Hasselmann, K., 1962: On the non-linear energy transfer in a gravity-wave spectrum. 1: General theory. *J. Fluid Mech*., **12**, 481.

Hasselmann, K., 1963a: On the non-linear energy transfer in a gravity-wave spectrum. 2: Conservation theorems, wave-particle correspondence, irreversibility. *J. Fluid Mech*. **15**, 273.

Hasselmann, K., 1963b: On the non-linear energy transfer in a gravity-wave spectrum. 3: Computation of the energy flux and swell-sea interaction for a Neumann spectrum. *J. Fluid Mech*. **15**, 385.

Hasselmann, S., Hasselmann, K., Allender, J.Z.H., Barnett, T.P., 1985: Computations and parameterizations of the nonlinear energy transfer for application in wave models. *J. Phys. Oceanogr.*, **15**, 1378-1391.

Hasselmann, K., Ross, D.B., Muller, P., and W. Sell, 1975: A parametric wave prediction model. *J. Physical Oceanogr.,* **6**, 200-228.

Hauser, D., Tison, C., Amiot, T., Delaye, L., & Corcoral, N., Castillan, P., 2017: SWIM: the First Spaceborne Wave Scatterometer. *IEEE Transactions on Geoscience and Remote Sensing*. in press. 10.1109/TGRS.2017.2658672.

Henderson, S. M., R. T. Guza, S. Elgar, and T. H. C. Herbers, 2006: Refraction of surface gravity waves by shear waves, *J. Phys. Oceanogr.*, **36**, 629–635, doi:10.1175/JPO2890.1.

Herbers, T.H.C., Elgar, S., Guza, R.T., 1994: Infragravity-Frequency (0.005-0.05 Hz) Motions on a Shelf. Part I: Forced Waves. *American Meteorological Soc.,* **24**, 917-927.

Herbers, T.H.C., Elgar, S., Guza, R.T., and W.C. O'Reilly, 1995(a): Infragravity-Frequency (0.005-0.05 Hz) Motions on a Shelf. Part II: Free Waves. *American Meteorological Soc.,* 1063-1079.

Herbers, T.H.C., Elgar, S., Guza, R.T., 1995(b): Generation and propagation of infragravity waves. *J. of Geophys. Res.*, **100**, C12, 24,863 – 24,872.

Herbers, T.H.C., and Burton, M.C., 1997. Nonlinear shoaling of directionally spread waves on a beach. *Journal of Geophysical Research,* **102**: doi: 10.1029/97JC01581.

Hersbach, H. and P.A.E.M. Janssen, 1999: Improvement of the short-fetch behavior in the wave ocean model (WAM). *J. Atm. Ocean. Techn.*, 16, 884-892.

Janssen, P. A. E. M., 1991: Quasi-linear theory of wind wave generation applied to wave forecasting. *J. Phys. Oceanogr.*, **21,** 1631–1642.

Janssen, P. A. E. M., K. Hasselmann, S. Hasselmann, and G. J. Komen, 1994: Parameterization of source terms and the energy balance in a growing wind sea. *Dynamics and Modelling of Ocean Waves,* G. J. Komen et al., Eds., Cambridge University Press, 215–238.

Janssen, T.T., J.A. Battjes and A.R. Van Dongeren, 2003: Long waves induced by short-wave groups over a sloping bottom. *J. Geoph. Res.*, **108**(C8), doi:10.1029/2002JC001515.

Janssen, T.T., Herbers, T.H.C., and J.A. Battjes, 2006: Generalized evolution equations for nonlinear surface gravity waves over two-dimensional topography, *J. Fluid Mech.* **552**, 393-418.

Janssen, T.T., T.H.C. Herbers and J. A. Battjes, 2008: Evolution of ocean wave statistics in shallow water: refraction and diffraction over seafloor topography. *J. Geoph. Res.*, **113**, doi:10.1029/2007JC004410.

Kaihatu, J.M., 2009: Application of a nonlinear frequency domain wave-current interaction model to shallow water recurrence effects in random waves. *J. Ocean Modelling*, **26,** 190-205.

Kaihatu, J., and Kirby, J., 1995: Nonlinear transformation of waves in finite water depth. *Physics of Fluids*, **7,** no. 8, 1903-1914.

Kaihatu, J. M.,Tahvildari, N., 2012: The combined effect of wave-current interaction and mud-induced damping on nonlinear wave. *Ocean Modelling*, **41**, 22–34. doi:10.1016/j.ocemod.2011.10.004.

Kobayashi, N., Raichle, A., and Asano, T., 1993: Wave Attenuation by Vegetation. *J. Waterway, Port, Coastal, Ocean Eng.*, 119:1(30), 30-48.  10.1061/(ASCE)0733-950X.

Komen, G.J., K. Hasselmann and S. Hasselmann, 1984: On the existence of a fully developed windsea spectrum. *J. Phys. Oceanogr.*, **14,** 1271-1285.

Leckler, F., Ardhuin, F., Filipot, J.F., and A. Mironov, 2013: Dissipation source terms and whitecap statistics.  *Ocean Mod.*, **70,** PP 62.74.

Leckler, F., Ardhuin, F., Peureux, C., Benetazzo, A., Bergamasco, F., and V. Dulov, 2015: Analysis and interpretation of frequency – wavenumber spectra of young wind waves.  *Am. Meteor. Soc.,* **45,** doi.org/10.1175/JPO-D-14-0237.1

Liu, Q., W. E. Rogers, A. V. Babanin and I. R. Young, L. Romero, S. Zieger, F. Qiao, C. Guan, submitted: Observation-based source terms in the third-generation wave model WAVEWATCH III: updates and verification, accepted for publication in *J. Phys. Oceanogr.*

Long, C.E., and Resio, D.T., 2007: Wind wave spectral observations in Currituck Sound, North Carolina, *J. Geophys. Res*. 112, CO5001.

Mendez F.J., Losada, I.J., 1999: Hydrodynamics induced by wind waves in a vegetation field. *J. of Geophys. Res.*, **104**, C8, 18383-18396.

Munk,W., 2009: An inconvenient sea truth: spread, steepness, and skewness of surface slopes. *Annual Review of Marine Sci*., **Vol 1**, 377-415.

Ng, C.N., 2000: Water waves over a muddy bed: a two-layer Stokes' boundary layer model. *Coast. Eng.* **40,** 221–242.

Perrie, W., and Resio, D.T., 2009: A Two-Scale Approximation for Efficient Representation of Nonlinear Energy Transfers in a Wind Wave Spectrum. Part II: Application to Observed Wave Spectra. *J. Physical Oceanography*. **39**, 2451–2476. DOI: 10.1175/2009JPO3947.1

Perrie, W., Toulany, B, Resio, D.T., Roland, A., and J.P. Auclair, 2013: A two-scale approximation for wave-wave interactions in an operational wave model, *Ocean Modell.*, **70**, 38 – 51.

Peureux, C., Benetazzo, A., and Ardhuin, F., 2018: Note on the directional properties of meter-scale gravity waves, *Ocean Sci.,* **14**, 41-52, https://doi.org/10.5194/os-14-41-2018.

Phillips, O. M., 1977: The dynamics of the upper ocean. Cambridge University Press, London, 336 p.

Phillips, O. M., 1984: On the response of short ocean wave components at a fixed wavenumber to ocean current variations. *J. Phys. Oceanogr.*, **14,** 1425–1433.

Phillips, 1985: Spectral and statistical properties of the equilibrium range in wind-generated gravity waves. *J. Fluid Mech.*, **156**, 505, doi:10.1017/S0022112085002221.

Pineau-Guillou, L., Ardhuin, F., Bouin, M.N., Redelsperger, J.L., Chapron, B., Bidlot, J.R., and Y. Quilfen, 2018: Strong winds in a coupled wave-atmosphere model during a North Atlantic storm event: evaluation against observations. *Royal Meteor. Soc.,* Vol **144**, Issue 711.

Pushkarev, A., Zakharov, V., 2016: Limited fetch revisited: Comparison of wind input terms, in surface wave modeling. *Ocean Modell.* **103**, 18–37.

Rascle, N. and Ardhuin, F., 2013: A global wave parameter database for geophysical applications. part 2: model validation with improved source term parameterization. *Ocean Modell.*, **70**, 174–188. doi:10.1016/j.ocemod.2012.12.001.

Resio, D.T., Long, C.E., Vincent, C.L., 2004: Equilibrium-range constant in wind-generated wave spectra. *J. Geophys. Res*. (Oceans) **109**, C01018.

Resio, D.T., Vincent, L., Ardağ, D., 2016: Characteristics of directional wave spectra and implications for detailed-balance wave modeling. *Ocean Modell.* **103** (2016), 38-52.

Resio, D.T. and W. Perrie, 2008: A two-scale approximation for nonlinear energy fluxes in a wind wave spectrum, Part I. *J. Phys. Oceanogr.* **38** (11), 2801-2816.

Rogers, W. E., P. A. Hwang, and D. W. Wang, 2003: Investigation of wave growth and decay in the SWAN model: Three regional-scale applications. *J. Phys. Oceanogr.*, **33,** 366–389.

Rogers, W.E. and D.W. Wang, 2006: On validation of directional wave predictions: review and discussion. Published by Naval Res. Lab., NRL/MR/7320- -06-8970.

Rogers, W. E. and K. T. Holland, 2009: A study of dissipation of wind-waves by mud at Cassino Beach, Brazil: Prediction and inversion. *Coastal Shelf Res.*, **29(3)**, 676–690.

Rogers, W.E., Babanin, A.V., Wang, D.W., 2012: Observation-consistent input and white-capping dissipation in a model for wind-generated surface waves: description and simple calculations. *J. Atmos. Oceanic Technol.* **29**(9), 1329-1346.

Rogers, W. E., M. D. Orzech, 2013: Implementation and testing of ice and mud source functions in WAVEWATCH III®. NRL Memorandum Report, NRL/MR/7320-13-9462, 31pp. (http://www7320.nrlssc.navy.mil/pubs.php)

Rogers, W.E., vanVledder, G.P., 2013: Frequency width in predictions of windsea spectra and the role of the nonlinear solver. *Ocean Modell.* **70**:52–61.

Roland, A. and Ardhuin, F. 2014: On the developments of spectral wave models: numerics and parameterizations for the coastal ocean. *Ocean Dynamics,* **64(6),** pp 833-846*.*

Romero. L., and W.K. Melville, 2010: Numerical Modeling of Fetch-Limited Waves in the Gulf of Tehuantepec. *J. Phys. Oceanogr.*, **40**, 466–486.

Romero, L., Melville, W.K. and J. Kleiss, 2012: Spectral energy dissipation due to surface-wave breaking. *J. Phys. Oceanogr.*, **42**, 1421–1444.

Ruessink, B.G., 1998: Bound and free infragravity waves in the nearshore zone under breaking and nonbreaking conditions, *J. Geophys. Res.*, **103**(C6), 12795–12805, doi:10.1029/98JC00893.

Saha, S., Moorthi, S., Pan, H.L., Wu, X., Wang, J., Nadiga, S., Tripp, P., Kistler, R., Woollen, J., Behringer, D., Liu, H., Stokes, D., Grumbine, R., Gayno, G., Wang, J., Hou, Y.T., ya Chuang, H., Juang, H.M.H.a.J.S., Iredell, M., Treadon, R., Kleist, D., Delst, P.V., Keyser, D., Derber, J., Ek, M., Meng, J., Wei, H., Yang, R., Lord, S., van den Dool, H., Kumar, A., Wang, W., Long, C., Chelliah, M., Xue, Y., Huang, B., Schemm, J.K., Ebisuzaki, W., Lin, R., Xie, P., Chen, M., Zhou, S., Higgins, W., Zou, C.Z., Liu, Q., Chen, Y., Han, Y., Cucurull, L., Reynolds, R.W., Rutledge, G., Goldberg, M., 2010: The NCEP Climate Forecast System Reanalysis. *Bull. Amer. Meterol. Soc*. **91,** 1015–1057.

Salmon J.E., Holthuijsen L.H., Zijlema M, van Vledder G.P., Pietrzak J.D., 2015: Scaling depth-induced wave-breaking in two-dimensional spectral wave models. Ocean *Model* **87**, 30–47. doi:10.1016/j.ocemod.2014.12.011.

Salmon J.E., Smit P.B., Janssen T.T., Holthuijsen L.H., 2016: A consistent collinear triad approximation for operational wave models. *Ocean Modelling,* **104**, 203-212.

Sheremet, A., Guza, R.T., Elgar, S. and T.H.C. Herbers, 2002: Observations of nearshore infragravity waves: Seaward and shoreward propagating components, *J. Geophys. Res.* **107**/C8, 3095, 10.1029/2001JC000970.

Sheremet, A., Guza, R.T. and T. H. C. Herbers, 2005: A new estimator for directional properties of nearshore waves J. *Geophys. Res.* **110**, C01001, doi:10.1029/2003JC002236.

Sheremet, A., Davis, J.R., Tian, M., Hanson, J., Hathaway, K., 2016: TRIADS: A phase-resolving model for nonlinear shoaling of directional wave spectra. *Ocean Modelling,* **99**, 60-74.

Smit P. and T. T. Janssen, 2013: The evolution of inhomogeneous wave statistics through a variable medium, *J. Phys. Ocean.*, **43**, 1741-1758.

Smit, P., Janssen, T.T., Holthuijsen, L., and J. Smith, 2014: Non-hydrostatic modeling of surf zone wave dynamics. *Coastal Engineering,* **83**, 36-48.

Smit, P. B., Janssen, T.T., and T. H. C. Herbers, 2015a: Stochastic modeling of coherent wave fields over variable depth. *J. Phys. Oceanogr.*, **45**, 1139–1154, doi:10.1175/JPO-D-14-0219.1.

Smit, P. B., Janssen, T.T. and T. H. C. Herbers, 2015b: Stochastic modeling of inhomogeneous ocean waves. *Ocean Modell.*, **96**, 26–35, doi:10.1016/j.ocemod.2015.06.009.

Smit, P.B. and T.T. Janssen, 2016: The evolution of nonlinear wave statistics through a variable medium, *J. Phys. Ocean.* **46**, 621-634.

Smith, G, Babanin, A.V., Riedel, P., Young, I.R., Oliver, S., and G. Hubbert, 2011: Introduction of a new friction routine into the SWAN model that evaluates roughness due to bedform and sediment size changes. *Coastal Engineering*, **58**, 317-326.

Stopa, J. E., Ardhuin, F., Husson, R., Jiang, H., Chapron, B., and Collard, F., 2016: Swell dissipation from 10 years of Envisat ASAR in wave mode. *Geophys. Res. Lett.*, **43**.

Stopa, J.E., Ardhuin, F., Babanin, A., Zieger, S., 2016a: Comparison and validation of physical wave parameterizations in spectral wave models. *Ocean Modell.,* **103**, 2-17.

Stopa, J. E., F. Ardhuin, and F. Girard-Ardhuin, 2016b: Wave climate in the Arctic 1992–2014: Seasonality and trends. *Cryosphere*, **10**, 1605–1629, doi: https://doi.org/10.5194/tc-10-1605-2016.

Sutherland, P., and W. K. Melville, 2015: Field measurements of surface and near-surface turbulence in the presence of breaking waves. *J. Phys. Oceanogr.*, **45**, 943–965.

Suzuki, T., Verwaest, T., Veale, W., Trouw, K. and Zijlema, M., 2012: A numerical study on the effect of beach nourishment on wave overtopping in shallow foreshores, in: P.J. Lynett and J.M. Smith (Eds.), Proc. 33th Int. Conf. on Coast. Engng., ASCE, World Scientific Publishing, Singapore, paper no. waves.50

The SWAMP Group, 1985: Sea Wave Modelling Project (SWAMP). An intercomparison study of wind wave prediction models, Part I: Principal results and conclusions. *Ocean Wave Modeling, Plenum Press,* pp 256.

Thomson, J., Ackley, S., Girard-Ardhuin, F, Ardhuin, F., Babanin, A, Boutin, G., Brozena, J., Cheng, S., Collins, C., Doble, M., Fairall, C., Guest, P., Gebhardt, C., Gemmrich, J., Graber, H., Holt, B., Lehner, S., Lund, B., Meylan, M., Wadhams, P., 2018: Overview of the Arctic Sea State and Boundary Layer Physics Program. *Journal of Geophysical Research*: Oceans. 10.1002/2018jc013766.

Toba, Y., 1972: Local balance in the air-sea boundary processes, I, On the growth process of wind waves, *J. Oceanogr. Soc. Jpn.*, **28**, 109-121.

Tolman, H. L., 1992: Effects of numerics on the physics in a third-generation wind-wave model. *J. Phys. Oceanogr.*, **22,** 1095–1111.

Tolman, H. L., and D. Chalikov, 1996: Source terms in a third-generation wind-wave model. *J. Phys. Oceanogr*, **26**, 2497-2518.

Tolman, H.L., 1998: Validation of NCEP's ocean winds for the use in wind wave models. *Global Atmos. Ocean Sys.* **6,** 243–268.

Tolman, H.L., 1998: Validation of a new global wave forecast system at NCEP. In: Edge, B.L., Helmsley, J.M. (Eds.), *Ocean Wave Measurements and Analysis*. ASCE, pp. 777–786.

Tolman, H. L., 2002: Validation of WAVEWATCH-III version 1.15. Tech. Rep. 213, NOAA/NWS/NCEP/MMAB.

Tolman, H.L., 2013: A generalized multiple discrete interaction approximation for resonant four-wave interactions in wind wave models. *Ocean Modell.* **70**, 11-24.

Tolman HL, Banner ML, Kaihatu JM, 2013: The NOPP operational wave model improvement project. *Ocean Modelling,* **70**, 2–10.

Tournadre, J., K. Whitmer, and F. Girard-Ardhuin, 2008: Iceberg detection in open water by altimeter waveform analysis. *J. Geophys. Res.*, **113,** C08040. doi:10.1029/2007JC004587.

Tournadre J., Girard-Ardhuin, F., Legrésy, B., 2012: Antarctic icebergs distributions, 2002- 2010. *Journal of Geophysical Research*, **117**, C05004, doi:10.1029/2011JC007441.

Tracy, B., and D. T. Resio, 1982: Theory and Calculation of the nonlinear energy transfer between sea waves in deep water. *Wave information study by the U.S. Army, Chief of Engineers*.

Tsagareli, K.N., Babanin, A.V., Walker, D.J., Young, I.R., 2010: Numerical investigation of spectral evolution of wind wave – Part I: Wind input function. *J. Phys. Oceanogr.* **40**, 656-666.

Van Vledder, G., 1999: Source term investigation: SWAN. Rev. 2, Rep. A162R1r2, Alkyon, 83 pp. plus figures.

Venkattaramanan, A. 2014: Nonlinear Characteristics of Wave Propagation over Vegetation. Master's thesis, Texas A & M University. Available electronically from http://hdl.handle.net/1969.1/152746.

WAMDI Group, (13 authors), 1988: The WAM model- a third generation ocean wave prediction model. *J. Phys. Oceanogr.* **18**, 1775-1810.

Wang, D. W., and P. A. Hwang, 2001: An Operational Method for Separating Wind Sea and Swell from Ocean Wave Spectra*. *J. Atmos. Ocean. Technol.*, **18**, 2052–2062, doi:10.1175/1520-0426(2001)018<2052:AOMFSW>2.0.CO;2.

WAVEWATCH III® Development Group (WW3DG), 2016: User manual and system documentation of WAVEWATCH III® version 5.16. *Tech. Note 329, NOAA/NWS/NCEP/MMAB, College Park, MD, USA*, 326 pp. + Appendices

Webb, D. J., 1978: Non-linear transfers between sea waves. *Deep Sea Res.*, **25**, 279–298, doi:10.1016/0146-6291(78)90593-3.

Young I.R. and A.V. Babanin, 2006: Spectral distribution of energy dissipation of wind-generated waves due to dominant wave breaking. *J. Phys. Oceanogr.*, **36**, 376-394.

Young, I. R., Babanin, A.V., and S. Zieger, 2013: The Decay Rate of Ocean Swell Observed by Altimeter. *J. Phys. Oceanogr.*, **43**, 2322–2333, doi:10.1175/JPO-D-13-083.1.

Zakharov, V.E., 1968: Stability of periodic waves of finite amplitude on the surface of deep water. *J. Appl. Mech. Tech. Phys*. **9**, 190-194.

Zhao, Z.D., Lian, J.J. and J.Z. Shi. 2006: Interactions among waves, current, and mud: numerical and laboratory studies, *Advances in Water Resources*, **29**, 1731–1744.

Zheng, C., Zhou, L., Shi, W., et al., 2015: Decadal variability of global ocean significant wave height. *J. Ocean Univ. China*, **14:** 778. https://doi.org/10.1007/s11802-015-2484-5.

Zieger S, Babanin AV, Rogers WE, Young IR, 2015: Observation-based source terms in the third-generation wave model WAVEWATCH. Virtual Special Issue on Ocean Surface Waves. *Ocean Model,* **96**, 2–25.

Table 1

Focus and Achievement of Depth Influenced Model Research Teams

| Working Group | Model | Study Focus | Algorithm Advancements |
|---|---|---|---|
| ST6 WG | SWAN | Bottom Friction | Coupled wave-bottom source term including flat bed and ripples formation/decay based on sediment size (Smith et al. 2011) |
| ST4 WG | WW3 | Bottom Friction, Depth induced breaking, currents, Triangular meshes | Use of triangle-based meshes- validation in strong tidal currents (Ardhuin et al. 2012) - coastal reflection<br>- ripple-induced bottom friction (SHOWEX parameterization)<br>- coupled with ocean circulation models using seamless deep to shallow breaking parameterization (Filipot & Ardhuin 2012)<br>-Post-processing to get bound and infra-gravity waves |
| Coherent wave propagation Janssen, Herbers, van Vledder, & Smit | SWAN & Research models | Inhomogeneity and nonlinear dynamics | -Coupled-mode transport generalization of the radiative transfer equation valid in coastal caustics (Smit & Janssen 2013; Smit, Janssen & Herbers 2015)<br>- Developed bi-Radiative Transfer Equation for calculation of bi-spectrum permitting estimates of harmonic and infra-gravity energy during shoaling<br>(Smit & Janssen 2016; Smit, Janssen, Holthuijsen, and Smith, 2014) |
| TRIADS Kaihatu, Sheremet, Smith and Hendrik Tolman | Separate | Nonlinear Shoaling, Mud, Vegetation | TRIADS –phase resolved hyperbolic resonant triad shoaling model<br>companion boundary layer model with muddy layer physics |
| Delft van Vledder, Holthuijsen | SWAN | Nonlinear Shoaling, depth induced breaking, flexible grid mesh | -Included improved high wind drag relationship in SWAN<br>-Deployed high resolution flexible mesh computations<br>-Improved parameterization of depth limited breaking<br>- Revisions of Stochastic Parametric Boussinesq and Lumped Triad Approximation for nonlinear shoaling effects on spectra |

Table 2

| Detailed description | ST2 | ST4 | ST5 | ST6 |
|---|---|---|---|---|
| Wind Input Term | Tolman and Chalikov(1996) based on distribution of momentum input as limited by the stress law formulation and impact on log-profile | From Ardhuin et al. (2010). Modification of Janssen(1991) -by including a wave supported stress (including both resolved and unresolved spectral components) adjustment to the log profile an-damping of input for very high frequencies and wind speeds | Modification of Janssen(1991). Input a function of wave age reflecting effect of waves on boundary layer and momentum transfer. Modification is a change in the angular dependence of the input. The extra wind stress due to separated air flow over breaking waves | Based on Donelan et al.(2005, 2006), Babanin et al. (2007), and Tsagareli et al. (2010). -Input adjusted to slow at strong winds (drag saturation) -Input consistent with total stress Negative Input for oblique or opposing winds Donelan (1999) and Aijaz et al. (2016) |
| Nonlinear Source Term | DIA | DIA | WRT FBI | DIA |
| Dissipation Source /Sink Terms | Tolman and Chalikov(1996). Two part function -a low frequency component analogous to energy dissipation via turbulence, -a high frequency component adjusted to the Phillips equilibrium range - both scaled to a wave age dependent energy level -empirical coefficients determined to weight the influence of each and to balance with the atmospheric input term | Ardhuin et al.(2010) - has saturation threshold component similar to Alves & Banner(2003) including isotropic and directional components - has a cumulative breaking element represents the smoothing of the surface of short waves by longer breaking waves that can be tied to the statistics of the breaking probability and energy loss of individual waves as in Banner et al.(2000), allowing estimation of the spectral density of breaking crest length per unit area. -swell dissipation due to effect of viscous and turbulent over the swell | Based on Alves&Banner(2003), Banner & Morison (2010). Has local and nonlocal components with the local component dependent upon local exceedance of saturation and the nonlocal dissipation related to the overall steepness of the waves. -active to reduce swell directly coupled to equations for breaking probability and strength - yields estimates of breaking wave crest statistics - also coupled to Sea Spray Droplet formulation modified from that of Fairall et al. (2009) - accounts for breaking crest wave speed slowdown (Banner et al. 2014) | Based on Babanin & Young (2005); Young & Babanin (2006), Babanin et al. (2010, Rogers et al. (2012) -threshold in terms of spectral density related to Phillips(1984) spectrum - first effect accounts for local wave breaking at a frequency number due to instability - second accounts for breaking induced by turbulence from longer breaking waves, but net effect is cumulative with respect to first -Non-breaking (swell) dissipation (Babanin 2011, 2012; Young et al. 2013, Zieger et al. 2015) based on approach of Ardhuin et al. (2009) accounting for effect of wind and whether airflow is viscous or turbulent. - dissipation constrained not to be more than wind input (Babanin et al. 2010) |
| Examples of Testing | Chawla et al. 2013 | Ardhuin et al. (2010) | Theoretical and lab tests winds 6-60 m/s, field data Straits of Juan De Fuca | Global Hindcast 2006, Tropical Cyclone Yasi |
| Available | WW3, SWAN | WW3, SWAN | WW3 | WW3, SWAN |

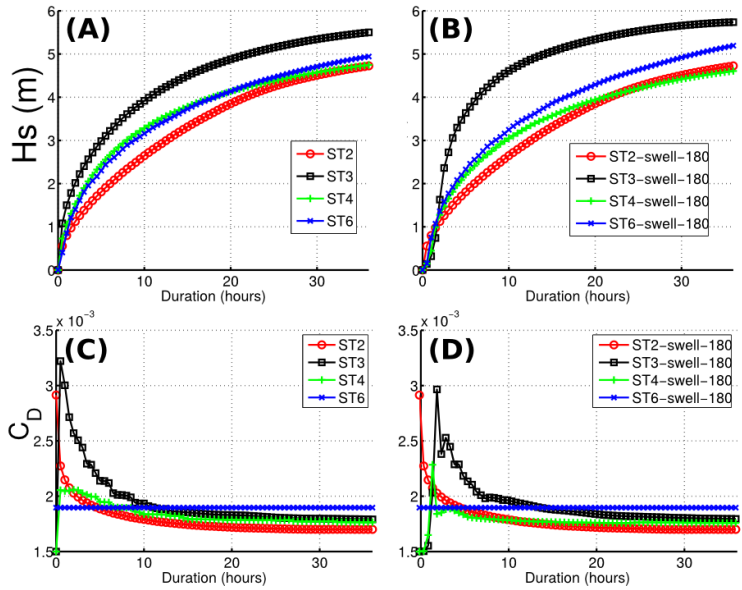

Figure 1: Illustration of the behavior or different parameterizations showing time-limited growth with a wind speed of 15 m/s starting either from rest (A and C) or from a 3 m swell of 18 s period. The top row shows the windsea wave height, and the bottom row shows the drag coefficient.

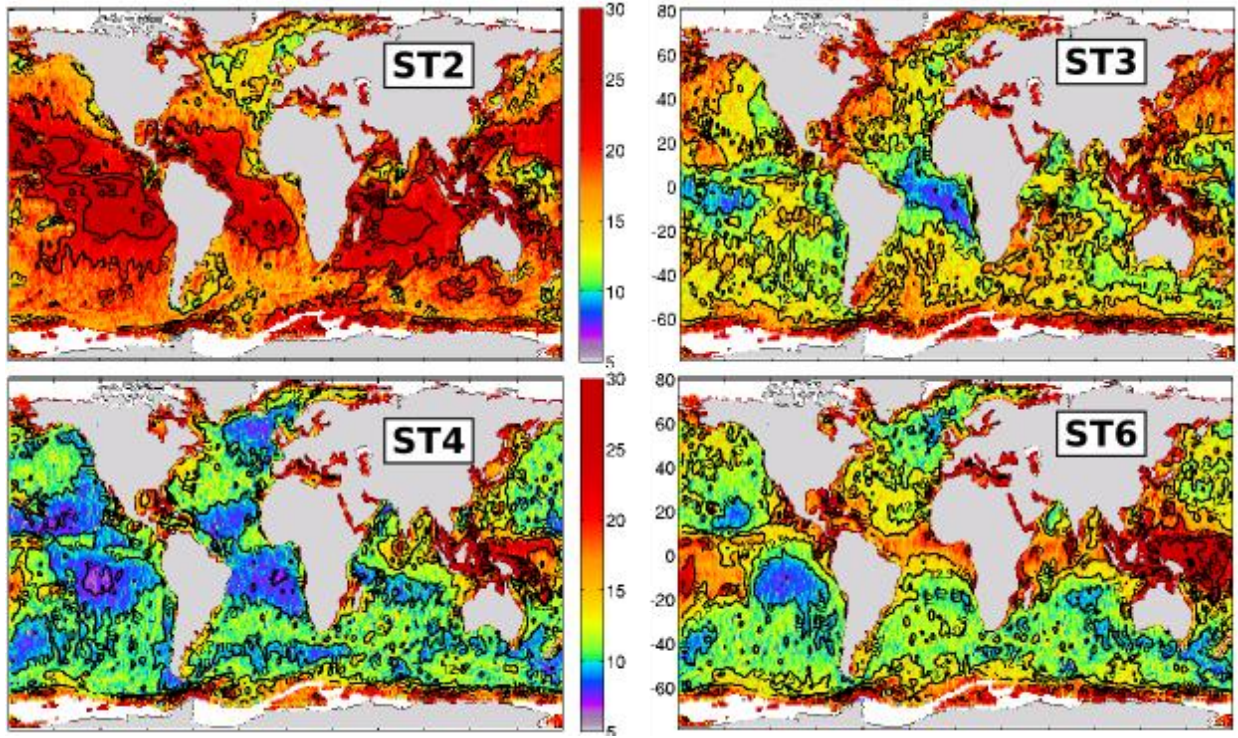

Figure 2: normalized root mean-square difference (in %) between all altimeters for the year 2011 (using 100 km along-track averages) and co-located model results obtained with ST2, ST3, ST4, and ST6.

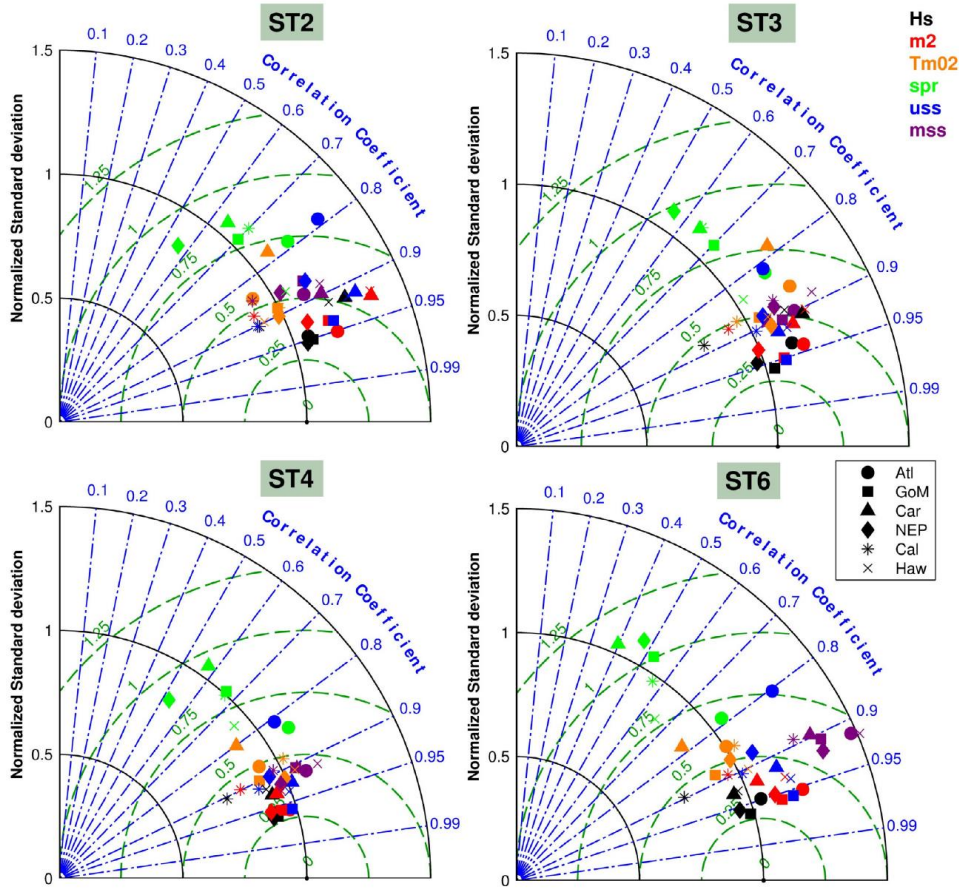

Figure 3: Taylor diagram – colors represent wave parameters significant wave height (black), orbital wave velocity at the surface (red), mean wave period (orange), average directional spread (green), Stokes surface velocity (blue), and the mean squared slope (purple). The six different symbols denote the regions: Atlantic (circle), Gulf of Mexico (square), Caribbean Sea (triangle), Northeast Pacific (diamond), California coast (star), and Hawaii (×). The different grid axis are the NSTD in solid black circles, the CRMSE in dashed green circles, and the correlation coefficient in blue dashed-dotted lines. (Reproduced from Stopa et al. 2016).

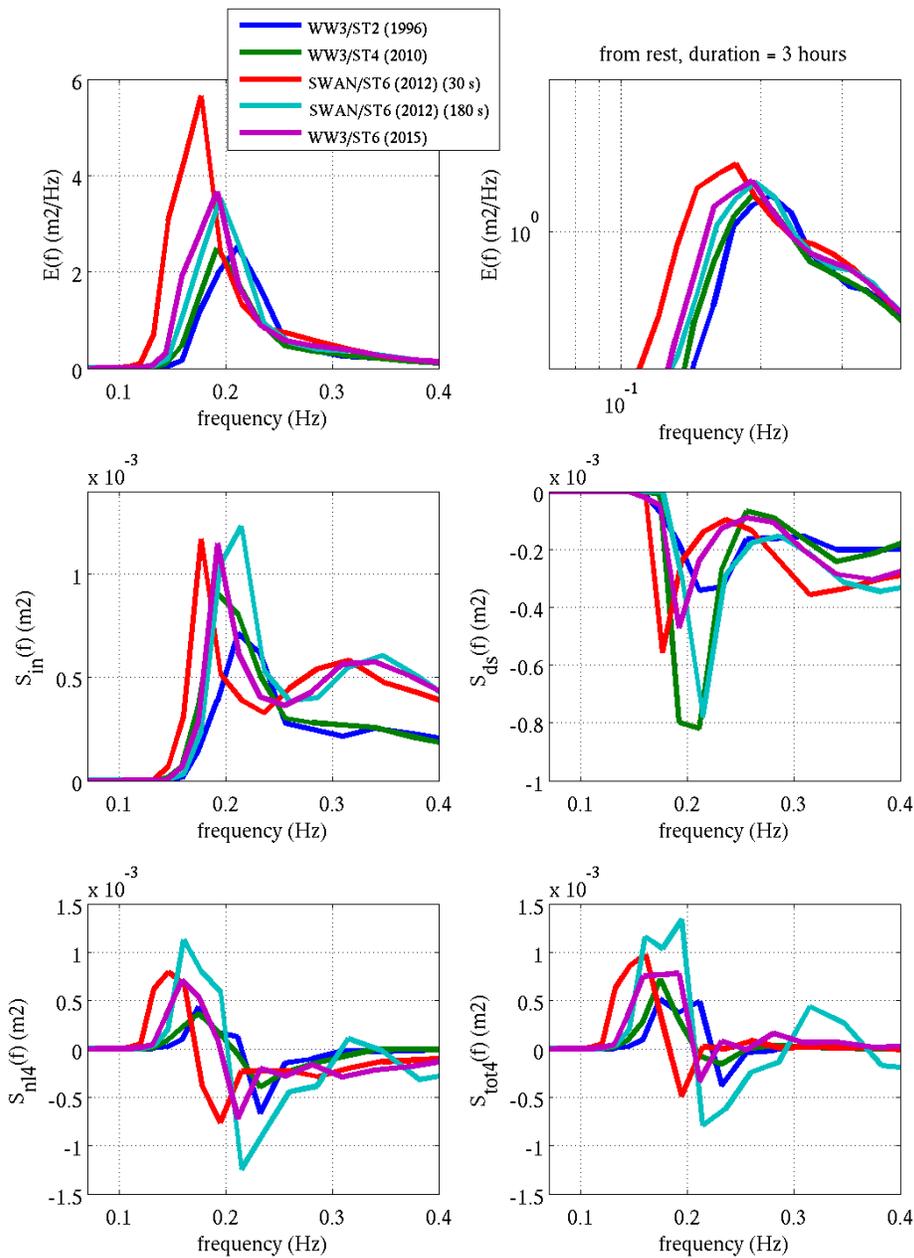

Figure 4. Directionally integrated source terms after 5 hours of simulation time for source terms packages as labeled with the top two panels showing the directional integrated energy spectrum on linear axes in the left panel and the right panel showing the same information on logarithmic axes. The second row shows the directionally integrated wind input on the left and the directionally integrated dissipation source term on the right side. The third row shows the directionally integrated nonlinear interaction source term on the left side and the total of all source terms on the right side.

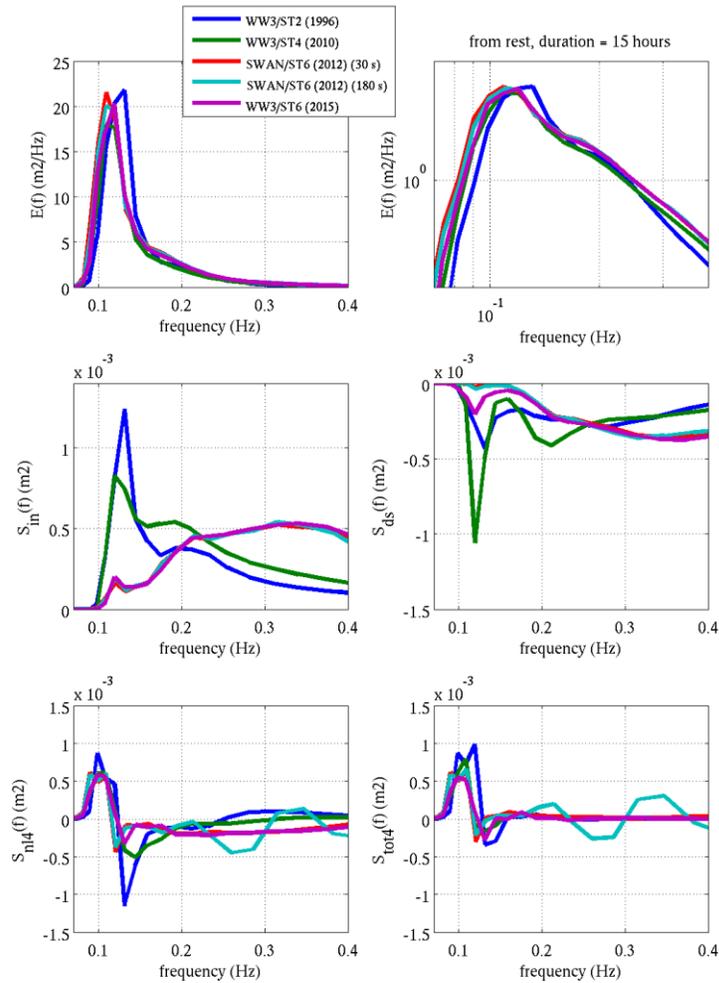

Figure 5. Directionally integrated source terms after 15 hours of simulation time for source terms packages as labeled with the top two panels showing the directional integrated energy spectrum on linear axes in the left panel and the right panel showing the same information on logarithmic axes. The second row shows the directionally integrated wind input on the left and the directionally integrated dissipation source term on the right side. The third row shows the directionally integrated nonlinear interaction source term on the left side and the total of all source terms on the right side.

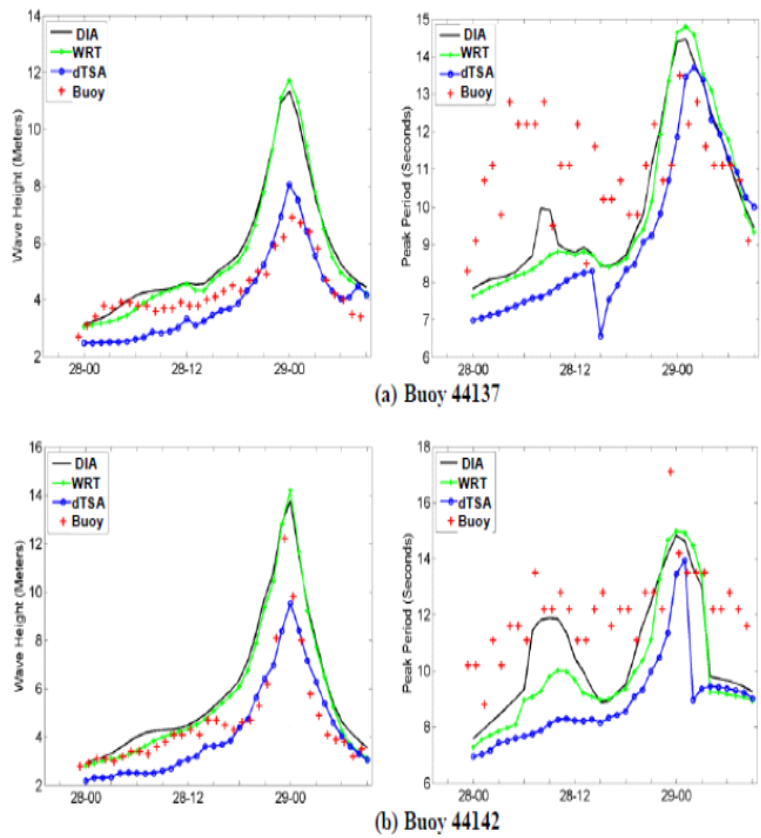

Figure 6. Comparison of improved TSA (dTSA) performance in operational applications at two deep-water buoys off the coast of Nova Scotia, based on an adaptation of the Perrie et al. (2013) version of the operational TSA.

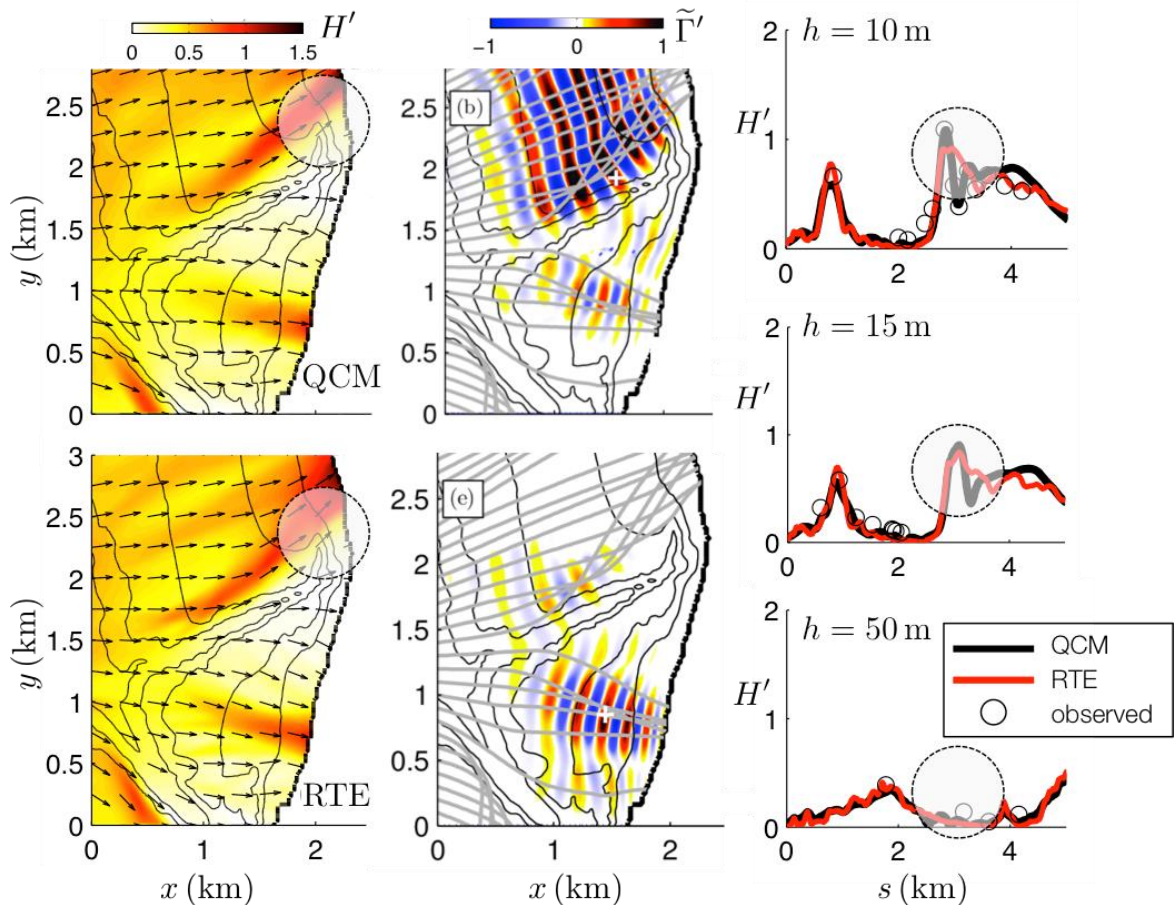

Figure 7. Comparison QC and RTE model simulations to observations of wave evolution across a nearshore canyon as observed during the ONR Nearshore Canyon Experiment (NCEX), in 2003 (see http://waveserver.org/nearshore-canyon-experiment-ncex/). The comparison is for a south swell field with a peak period of approximately 18s, as observed on 16 November 2003 [see Smit et al, 2015a]. Left panels show normalized wave heights for QC approximation (top) and RTE (bottom). Right panels show normalized wave height along 10m (top), 15m (middle) and 50m (bottom) depth contours. In most regions the agreement between observations and models is quite good and comparable. However, near the canyon heads (indicated with circles) where coherent effects are important, the wave heights predicted by the QC model are in much better agreement with the observations (where available) than the RTE model predictions. Middle panels show the spatial covariance function of the same wave field. Solid grey lines represent wave rays computed for this wave field. Top middle panel shows the covariance function $G(\mathbf{x}, \mathbf{x}\ell)$ with $\mathbf{x}\ell$ constant and located just north of the Scripps canyon head (indicated by white cross). Bottom middle panel shows the covariance function centered roughly midway between La Jolla and Scripps canyon (white cross). From the covariance function centered north of the canyon head (middle top panel) we see that a nodal pattern spreads out to the north and some correlation exists between waves north and south of the canyon. The covariance function centered between the canyon heads (middle bottom panel) is indicative of a fairly homogeneous wave field (for which the RTE would be a reasonable approximation). The spatial covariance function is accurately captured in the QC model approximation, providing insight in the spatial correlations in the wave field; this information is generally not available in the quasi-homogeneous approximation represented by the RTE.

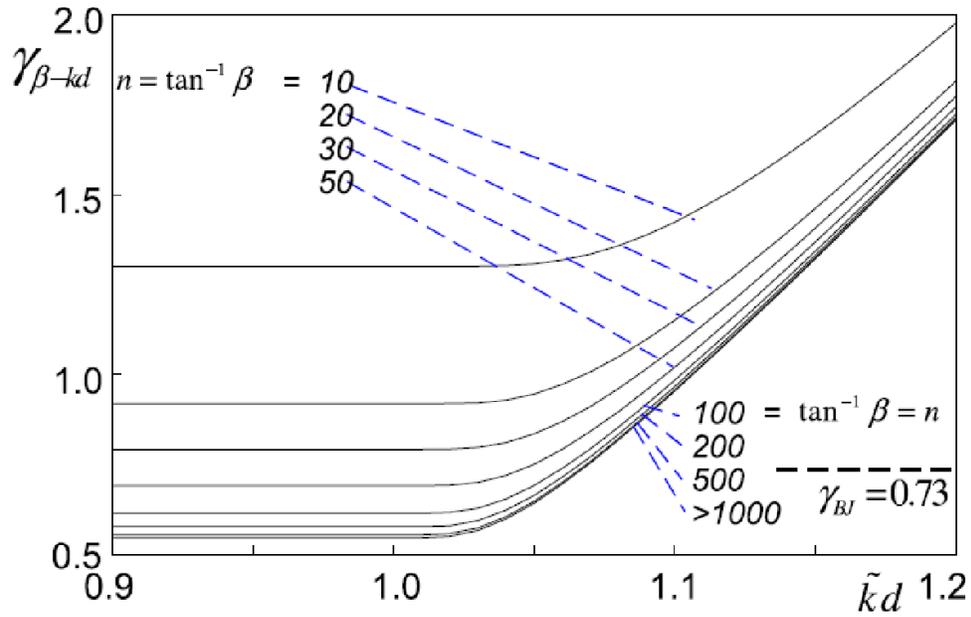

Figure 8. Update to the Battjes-Janssen Breaking Model. Calibrated $\gamma_{\beta-kd}$ as a function of bottom slope $n = \tan^{-1}\beta$ and normalized characteristic wave number $\tilde{k}d$ and $\gamma_{BJ} = 0.73$ for reference.

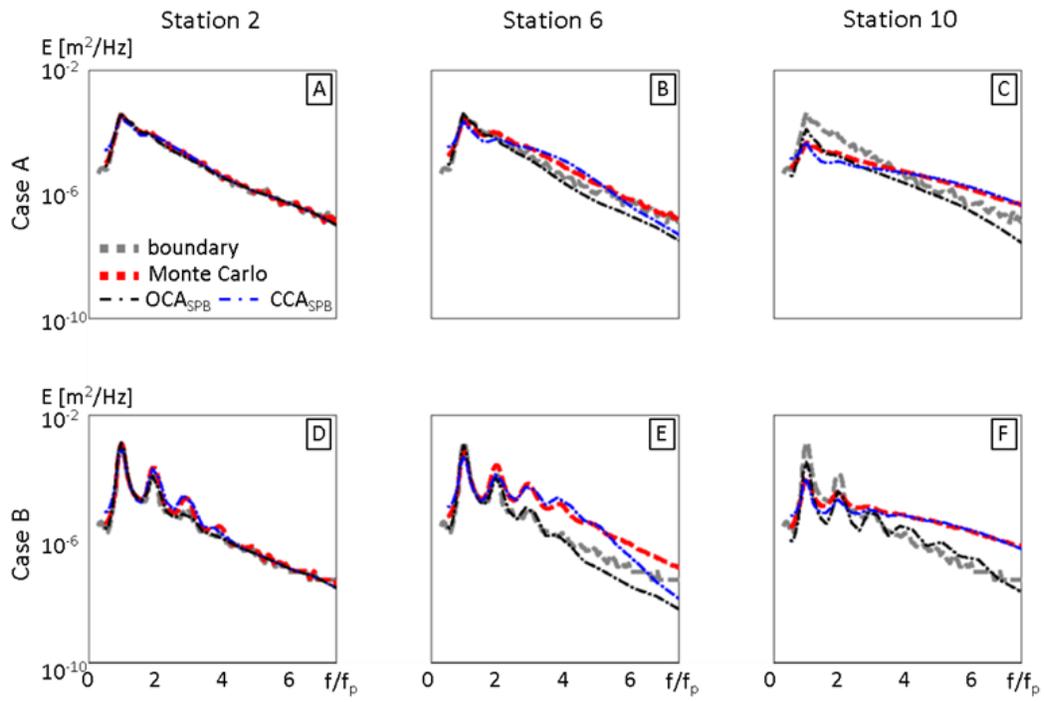

Figure 9. Simulation of Higher Order Harmonics in the SWAN Model with New SPB formulation.

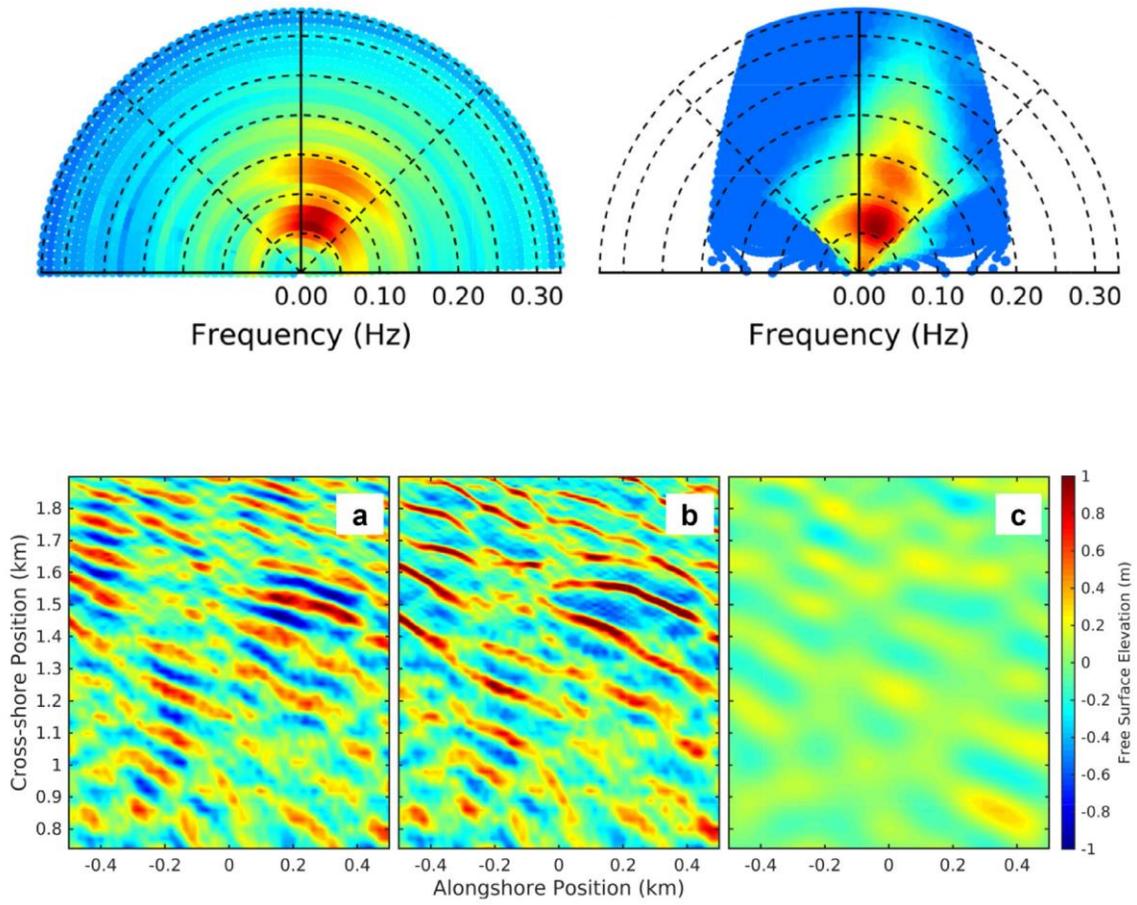

Figure 10. TRIAD Simulation of Hurricane Bill, at Duck NC.

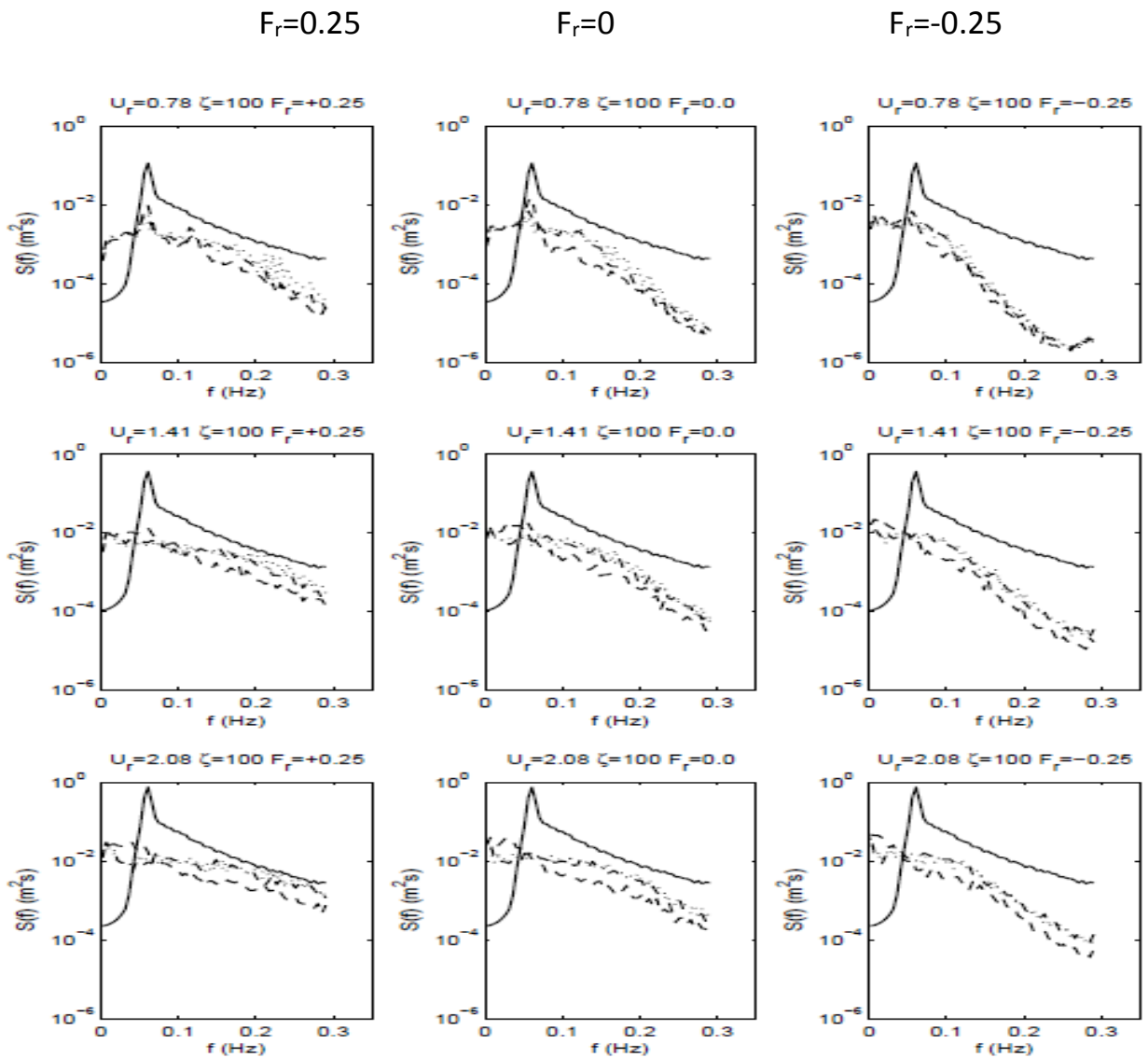

Figure 11. Wave-mud-current interaction over a flat bottom (depth of 2m) from a nonlinear spectral model: high degree of damping. Solid line is the initial spectrum, dashed-dot line is the spectrum a distance of 21 times the wavelength associated with the peak period away, and dashed line is a distance of 50 times the wavelength associated with the peak period away. Rows: Common Ursell numbers. Columns: Common Froude numbers associated with the velocity and direction of the flow.

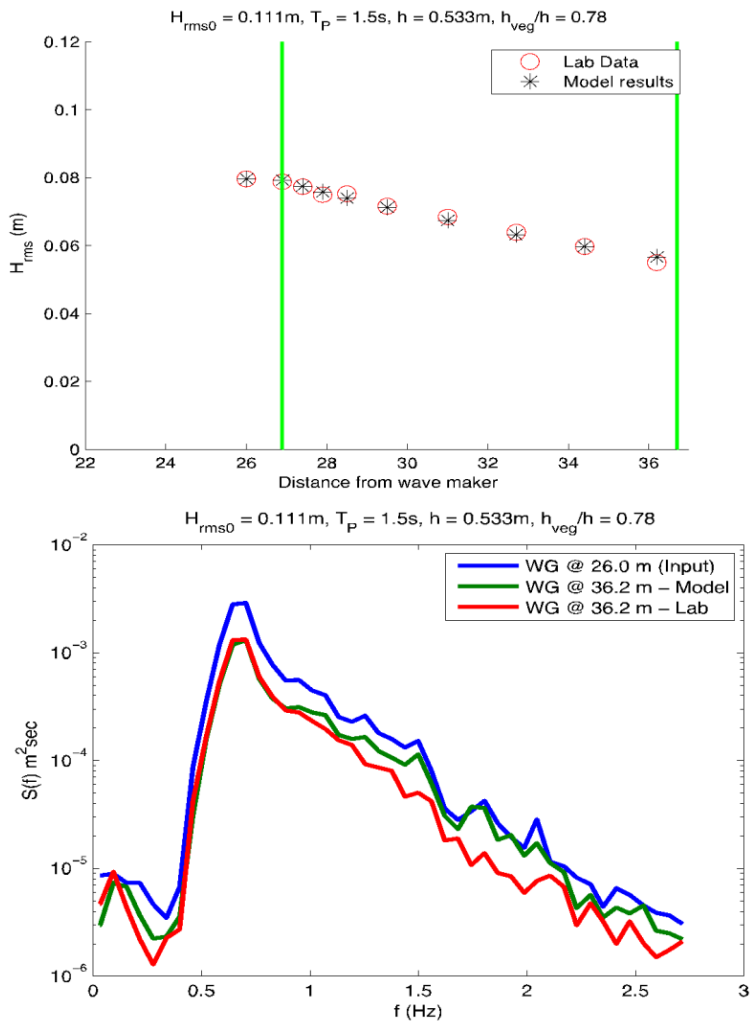

Figure 12. Comparison of one dimensional nonlinear wave and vegetation interaction model with laboratory data from Anderson and Smith (2013). Top panel compares $H_{rms}$ between model and laboratory at different distances of propagation through vegetation. Bottom panel provides comparison of the spectra.